\documentclass[11pt]{article}

\usepackage{amsmath,amsfonts,amsbsy,amssymb,array,accents,dsfont}
\usepackage{enumerate,array,latexsym,graphicx,mathrsfs,verbatim,psfrag}
\usepackage{bm} 
\usepackage[normalem]{ulem}
\usepackage{booktabs} 
\usepackage[usenames]{color}

\usepackage[english]{babel}
\usepackage{fancybox}

\usepackage{datetime}

\usepackage[nosort]{cite}
\usepackage{chngpage} 

\usepackage{enumitem}
\setlist{itemsep=0pt}

\usepackage[colorlinks=true,      linkcolor=darkblue,      urlcolor=darkblue,      
            filecolor=darkblue,      citecolor=darkblue,       pdfstartview=FitH,     
						pdfpagemode=UseNone,      bookmarksopen=true]{hyperref}  
\usepackage[all]{hypcap}     

\def\eq#1{(\ref{#1})}

\newcommand{\captionfonts}{\small}
\makeatletter  
\long\def\@makecaption#1#2{%
  \vskip\abovecaptionskip
  \sbox\@tempboxa{{\captionfonts #1: #2}}%
 \ifdim \wd\@tempboxa >\hsize
    {\captionfonts #1: #2\par}
  \else
    \hbox to\hsize{\hfil\box\@tempboxa\hfil}%
  \fi
  \vskip\belowcaptionskip}
\makeatother   

\DeclareMathSymbol{\medhatsym}{\mathord}{largesymbols}{"62} 
\newcommand\lowermedhatsym{
  \text{\smash{\raisebox{-1.28ex}{%
    $\medhatsym$}}}}
\newcommand\medhat[1]{
  \mathchoice
    {\accentset{\displaystyle\lowermedhatsym}{#1}}
    {\accentset{\textstyle\lowermedhatsym}{#1}}
    {\accentset{\scriptstyle\lowermedhatsym}{#1}}
    {\accentset{\scriptscriptstyle\lowermedhatsym}{#1}}
}

\DeclareMathSymbol{\medtildesym}{\mathord}{largesymbols}{"65}
\newcommand\lowermedtildesym{
  \text{\smash{\raisebox{-1.2ex}{%
    $\medtildesym$}}}}
\newcommand\medtilde[1]{
  \mathchoice
    {\accentset{\displaystyle\lowermedtildesym}{#1}}
    {\accentset{\textstyle\lowermedtildesym}{#1}}
    {\accentset{\scriptstyle\lowermedtildesym}{#1}}
    {\accentset{\scriptscriptstyle\lowermedtildesym}{#1}}
}

\newcommand{\comm}[1]{} 

\def\IQ{\mathbb{Q}}
\def\IR{\mathbb{R}}
\def\IS{\mathbb{S}}
\def\IT{\mathbb{T}}
\def\IZ{\mathbb{Z}}

\def\({\left(}
\def\){\right)}
\def\[{\left[}
\def\]{\right]}

\def\gstr{g_{\rm s}}

\def\sst{\scriptscriptstyle}

\def\coeff#1#2{{\textstyle \frac{#1}{#2}}}

\def\One{{\hbox{ 1\kern-.8mm l}}}

\def\barray{\begin{array}}
\def\earray{\end{array}}
\def\be{\begin{equation}}
\def\ee{\end{equation}}
\def\bea{\begin{eqnarray}}
\def\eea{\end{eqnarray}}
\def\bal{\begin{align}}
\def\eal{\end{align}}
\def\nn{\nonumber}

\def\hlat{h^\#}

\def\mR{\mathds{R}}
\def\mZ{\mathds{Z}}

\numberwithin{equation}{section} 

\makeatletter
\g@addto@macro\bfseries{\boldmath}
\makeatother

\definecolor{cardinal}{rgb}{0.6,0,0}
\definecolor{darkgreen}{rgb}{0,0.4,0}
\definecolor{purple}{rgb}{0.5, 0, 0.5}
\definecolor{golden}{rgb}{0.92, 0.7, 0}
\definecolor{midnight}{rgb}{0, 0, 0.5}
\definecolor{darkblue}{rgb}{0, 0, 0.8}

\def\Neql#1{{\cal N}\!=\!{#1}}
\def\coeff#1#2{\relax{\textstyle {#1 \over #2}}\displaystyle}

\def\QQ{\mathds{Q}}
\def\IR{\mathds{R}}
\def\ZZ{\mathds{Z}}

\def\cB{{\cal B}}

\def\cF{{\cal F}}

\def\cH{{\cal H}}

\def\cM{{\cal M}}
\def\cN{{\cal N}}

\def\cP{{\cal P}}
\def\cQ{{\cal Q}}
\def\cR{{\cal R}}

\def\cW{{\cal W}}

\def\nBPS#1{$\frac{1}{#1}$-BPS}

\def\bbS{\mathbb{S}}

\def\sst#1{\scriptscriptstyle{#1}}

\newcommand{\nk}{\ensuremath{\kappa}}

\newcommand{\lambdafour}{\lambda_{4}}
\newcommand{\lambdahatfour}{\hat\lambda_{4}}

\newcommand{\thetafour}{\Theta_{4}}
\newcommand{\thetahatfour}{\medhat\Theta_{4}}

\newcommand{\nv}{\ensuremath{N_{\mathrm{\sst{V}}}}}
\newcommand{\nc}{\ensuremath{N_{\mathrm{\sst{C}}}}}

\newcommand{\T}[3]{\ensuremath{ #1{}^{#2}_{\phantom{#2} \! #3}}}		

\newcommand{\subsubsectionmod}[1]{
\refstepcounter{subsubsection}
\subsubsection*{\thesubsubsection ~~  #1}
}

\newcommand{\subsectionmod}[1]{
\refstepcounter{subsection}
\subsection*{\thesubsection ~~  #1}
}

\begin{document}


\begin{flushright}
IPHT-T17/051\\
\end{flushright}

\vspace{15mm}

\begin{center}

{\huge \bf{M-theory Superstrata and the MSW String}}

\vspace{14mm}

\bigskip\bigskip
\centerline{{\bf Iosif Bena$^1$, Emil Martinec$^{2}$, David Turton$^1$,  and Nicholas P. Warner$^{3}$}}
\bigskip
\bigskip

\centerline{$^1$ Institut de Physique Th\'eorique,}
\centerline{Universit\'e Paris Saclay, CEA, CNRS, }
\centerline{Orme des Merisiers,  F-91191 Gif sur Yvette, France}
\bigskip
\centerline{$^2$Enrico Fermi Inst. and Dept. of Physics, }
\centerline{University of Chicago,  5640 S. Ellis Ave., }
\centerline{Chicago, IL 60637-1433, USA}
\bigskip
\centerline{$^3$ Department of Physics and Astronomy,}
\centerline{and Department of Mathematics,}
\centerline{University of Southern California,} \centerline{Los
Angeles, CA 90089, USA}

\bigskip\bigskip\bigskip

 
\textsc{Abstract}

\begin{adjustwidth}{17mm}{17mm} 
 %
\vspace{4mm}
\noindent
The low-energy description of wrapped M5 branes in compactifications of M-theory on a Calabi-Yau threefold times a circle is given by a conformal field theory studied by Maldacena, Strominger and Witten and known as the MSW CFT. Taking the threefold to be $\IT^6$ or K3$\times \IT ^2$,
we construct a map between a sub-sector of this CFT and a sub-sector of the D1-D5 CFT. We demonstrate this map by considering a set of D1-D5 CFT states that have smooth horizonless bulk duals, and explicitly constructing the supergravity solutions dual to the corresponding states of the MSW CFT.  We thus obtain the largest known class of solutions dual to MSW CFT microstates, and demonstrate that five-dimensional ungauged supergravity admits much larger families of smooth horizonless solutions than previously known.
%
\end{adjustwidth}

\end{center}


\thispagestyle{empty}

\newpage


\baselineskip=14pt
\parskip=2pt

\tableofcontents

\newpage

\baselineskip=15pt
\parskip=3pt

\section{Introduction}
\label{Sect:introduction}

For around twenty years there have been two well-known routes to describe the entropy of BPS black holes in some form of weak-coupling limit.  The first of these was done using the original perturbative, weak-coupling microstate counting of momentum excitations of the D1-D5 system \cite{Strominger:1996sh}.   The corresponding three-charge black hole, in five dimensions, is obtained by compactifying  D1 and D5 branes of the IIB theory on $\cM \times \IS^1$, where $\cM$ is either $\IT^4$ or  K3, and adding momentum. When the size of the $\IS^1$ wrapped by $N_1$ D1 and $N_5$ D5 branes is the largest scale in the problem, the low-energy physics is given by a superconformal field theory (SCFT) with central charge $6 N_1 N_5$. There is now strong evidence that there is a locus in moduli space where this SCFT is a symmetric product orbifold theory with target space $\cM^N/S_N$, where $N=N_1 N_5$~\cite{Vafa:1995zh}. This orbifold point is naturally thought of as the weak-coupling limit on the field theory side, and it is far in moduli space from the region in which supergravity is weakly coupled (see for example the review~\cite{David:2002wn}).  

The D1-D5 system has eight supersymmetries that, in terms of left- and right-moving modes, are   $\cN=(4,4)$.  Adding $N_P$ units of left-moving momentum breaks the supersymmetry to $(0,4)$. The state counting emerges from the ways of partitioning this momentum amongst the fundamental excitations of the SCFT, and the result matches the entropy of the three-charge black hole in five dimensions. 

The D1-D5 black string in six dimensions has a near-horizon limit that is  AdS$_3 \times \IS^3$. The holographic duality between the strongly coupled D1-D5 CFT and supergravity on AdS$_3 \times \IS^3$ has been widely studied and is, as far as these things go, relatively well understood. In particular, one can often do ``weak-coupling'' calculations in the orbifold CFT that can be mapped to ``strong-coupling" physics described by the AdS$_3 \times \IS^3$ supergravity dual of this theory.  The computation of BPS black hole entropy is one such example. However the holographic duality gives much more information than simple entropy counts, enabling the study of strong-coupling physics of individual microstates of the CFT, and their bulk descriptions (see, for example, \cite{Lunin:2001jy,Giusto:2004id,Giusto:2004ip,Skenderis:2006ah,Kanitscheider:2006zf,Mathur:2011gz,Giusto:2012yz,Bena:2015bea,Giusto:2015dfa,Bena:2016agb,Bena:2016ypk}). 

The second approach is to use the Maldacena-Strominger-Witten (MSW)~\cite{Maldacena:1997de} ``string.''  Here one starts with a compactification of M-theory on a Calabi-Yau threefold to five dimensions.   One then wraps an M5 brane around a suitably chosen divisor to obtain a $(1+1)$-dimensional string in five dimensions. This breaks the supersymmetry to $\cN=(0,4)$ right from the outset. There is a $(1+1)$-dimensional SCFT on the worldvolume of this string and its central charge is proportional to the number of moduli of the wrapped brane.\footnote{The central charge can be computed in terms of the intersection properties of the divisor through a simple index theorem.} This MSW string is wrapped around another compactification circle to obtain a four-dimensional compact bound state.  One can add momentum excitations to the  MSW string in a manner that preserves the $(0,4)$ supersymmetry, and the entropy of these excitations matches (to leading order) the entropy of the corresponding four-dimensional BPS black hole.  

However, despite multiple attempts over the past twenty years, the holographic description of the MSW black hole remains much more mysterious. The strongly-coupled physics of this CFT is described by weakly-coupled supergravity in an asymptotically AdS$_3 \times \IS^2 \times {\rm CY}$ geometry, but it appears that there does not exist an exactly solvable, symmetric-orbifold CFT anywhere in the moduli space~\cite{Larsen:1999uk,Larsen:1999dh}.

The purpose of this paper is to construct a map between a large sub-sector of the MSW CFT and a large sub-sector of the D1-D5 CFT.  Our construction has two steps. We first construct a map between the Type IIB  [global AdS$_3]\times \IS^3 \times \IT^4$ solution dual to the NS vacuum of the D1-D5 CFT (and $\IZ_\kappa$ orbifolds thereof) and the M-theory [global AdS$_3] \times \IS^2 \times \IT^6$ solution dual to the NS vacuum of the MSW CFT.\footnote{More precisely, beginning with a set of D1-D5 R-R ground states that are related to ($\IZ_\kappa$ orbifolds of) [global AdS$_3]\times \IS^3 \times \IT^4$~\cite{Balasubramanian:2000rt,Maldacena:2000dr}, we construct a map to the MSW maximally-spinning Ramond ground state, which is related to [global AdS$_3]\times \IS^2 \times \IT^6$ by spectral flow.} The first step of this map converts the $\IZ_\kappa$ orbifold of the D1-D5 solutions into a smooth $\IZ_\kappa$ quotient on the Hopf fiber of the $\IS^3$, which adds (the near-horizon limit of) KK-monopole charge. The second step involves a known sequence of T-dualities and lift to M-theory. We then observe that this map also takes all the D1-D5 microstate geometries that are independent of the Hopf fiber into microstate geometries of the MSW system. 

Very large families of microstate geometries of the D1-D5 CFT, parameterized by arbitrary continuous functions of two (or perhaps even three) variables, have been constructed over the past few years using {\it superstratum} technology~\cite{Bena:2015bea,Bena:2016agb,Bena:2016ypk}.   In six dimensions, these solutions are expanded into three sets of Fourier modes, labelled by $(k,m,n)$.   Our map takes the modes with $k=2m$ to asymptotically AdS$_3\times \IS^2$ solutions dual to momentum-carrying microstates of the MSW CFT.   Since, in principle, D1-D5-P superstrata with generic $(k,m,n)$  are described by functions of three variables\footnote{Smooth solutions with generic Fourier modes have not yet been explicitly constructed, and it is conceivable that interactions between such generic modes could introduce new, unanticipated singularities.  This is why we are using the phrasing ``in principle.''}, the restriction to enable the map to MSW CFT reduces this to functions of two variables. In practice, in this paper we have constructed solutions with $k=2$, $m=1$ and generic $n$ and so our M-theory superstrata are parameterized by one integer $n$. Extrapolating to superpositions of two modes will give, by the linearity of the BPS equations of six-dimensional supergravity, smooth solutions parameterized by functions of one variable.\footnote{Another way to build M-theory superstrata parameterized by a function of one variable is to impose the $k=2m$ condition on the superstrata constructed in~\cite{Bena:2015bea}.} 
Either way, we are able to build the largest known class of smooth microstate geometries for the MSW black hole. 

Precise dual CFT states for superstrata solutions have been identified~\cite{Bena:2015bea,Giusto:2015dfa,Bena:2016agb,Bena:2016ypk}, and our map indicates that there is a one-to-one map between the subset of these states that are eigenstates of the $\cR$-current $J_L^3$ corresponding to rotations around the Hopf fiber of the $\IS^3$, and a certain class of states of the MSW CFT. Given that the MSW CFT does not seem to have a weakly-coupled symmetric orbifold description, the fact that one can map a sector of this CFT into a sector of the D1-D5 CFT may provide leverage in analyzing some aspects of the MSW CFT, such as the set of protected three-point functions where two of the operators are heavy and one is light.

The  [global AdS$_3] \times \IS^2 \times \IT^6$ solution dual to the NS vacuum of the MSW CFT can be obtained as an uplift of a Type IIA configuration with two fluxed D6 branes of opposite charges (the fluxes give rise to D4 charges which uplift to the M5 charges of the M-theory solution).  The geometric transition that employs fluxed  D6 (and anti-D6) branes to convert black holes and black rings  into smooth, horizonless geometries  was first described in \cite{Bena:2005va,Berglund:2005vb}. A particular example of this was studied in \cite{Denef:2007yt} in which a fluxed D6-$\overline{\rm D6}$  bound state was used as a background to study the dynamics of D0 branes and wrapped D2 branes.  This is sometimes referred to black-hole and black-ring ``deconstruction''  \cite{Denef:2007yt,Gimon:2007mha}.  

Adding D0 branes to the D6-$\overline{\rm D6}$ configuration corresponds, in the M-theory uplift, to adding momentum along the AdS$_3$ angular direction, and this configuration was studied in detail in several papers almost a decade ago. There are several ways in which this can be done.  First, one can add ``pure'' D0's, which are free to move on a hyperplane in the solution  \cite{Denef:2007yt,deBoer:2008zn,deBoer:2009un}. Uplifting this supergravity solution to M-theory gives rise to a singular supergravity PP-wave solution, carrying angular momentum along both AdS$_3$ and $\IS^2$; however, this naive extrapolation ignores the non-Abelian and non-linear dynamics of multiple D0-branes.
A second way to add momentum charge is to add a gas of supergravity modes directly in M-theory, in the smooth [global AdS$_3] \times \IS^2 \times \IT^6$ solution. The entropy of this ``supergraviton gas'' \cite{deBoer:1998us}  scales in the same way as the added D0 branes described above~\cite{deBoer:2009un}. However the back-reaction of this supergraviton gas was not constructed.

The solutions we find are smooth M-theory geometries carrying the same charges as the foregoing ensembles of states, the D0's and the supergraviton gas, and so it is natural to think of our solutions as examples of fully back-reacted smooth geometries associated to the supergraviton gas, or to correctly-uplifted D0 branes.  Indeed, the linearized limit of the superstratum modes is explicitly a supergraviton gas in [global AdS$_3] \times \IS^2 \times \IT^6$.  Hence, at least outside the black-hole regime of parameters, it may well be that some of our M-theory solutions are fully back-reacted supergraviton gas states.

In the black-hole regime of parameters, one desires more entropy than is provided by the supergraviton gas.  
There are other methods of incorporating D0-brane charge in this regime, which often go beyond supergravity. For example, one can place branes in the type IIA background that carry D0 charge as a worldvolume flux. One possibility is to add a D4-D2-D2-D0 center, which uplifts in M-theory to a M5-M2-M2-P supertube that rotates along the AdS$_3$~\cite{Bena:2010gg}.  Since supertubes can have arbitrary shapes~\cite{Mateos:2001qs}, the solutions corresponding to these configurations can have a non-trivial dependence on the M-theory circle; however, like the pure D0-brane sources, these are again naively singular configurations.   Estimates of the entropy of back-reacted solutions thus far yield results sub-leading relative to the black hole entropy~\cite{Bena:2010gg}.

Another possibility is to add D0 branes via world-volume flux on a dipolar, egg-shaped D2-brane\cite{Denef:2007yt}.  Counting the Landau levels of this two-brane has been argued to reproduce the BPS black hole entropy.  This ``egg-brane'' uplifts in M theory to an M2 brane wrapping the $\IS^2$ and spinning in AdS$_3$, and the corresponding solution is also singular \cite{Raeymaekers:2014bqa,Raeymaekers:2015sba}.  Moreover, simply wrapping branes around this $\IS^2$ adds yet another charge to the system, which either (a) introduces an uncanceled tadpole which changes the asymptotics of the supergravity fields; or (b) breaks all of the supersymmetry \cite{Tyukov:2016cbz}.  Either way, such configurations cannot represent BPS microstates of the original black hole. 

It is possible that smooth geometric oscillations of superstrata in deep scaling geometries might contribute a finite fraction of black hole entropy~\cite{Bena:2014qxa}, but this is by no means proven.   Such states lie well beyond the consideration of a supergraviton gas in the [global AdS$_3] \times \IS^2$ background.
Our construction gives new possibilities for deep superstrata in the M-theory frame, and thus represent another advance in the quest for a geometrical understanding of black hole entropy.
The fields that make smoothness of superstrata geometries possible are exactly of the kind one expects to see when one considers the back-reaction of the momentum-carrying M5 brane source in the MSW system, or that are generated in string emission calculations \cite{Giusto:2011fy} in a U-dual four-charge configuration of D3 branes \cite{Bianchi:2016bgx}\footnote{These fields are absent in the solutions of \cite{deBoer:2008zn,Raeymaekers:2014bqa,Raeymaekers:2015sba,Bena:2010gg}.}.   We believe that this is not a coincidence but rather an indication that our construction is closing in on a good holographic description of the microstates of this system.  

Besides its interest for understanding the MSW CFT, our map is also a very powerful solution-generating device. Indeed, we will use it to construct new smooth solutions of five-dimensional ungauged supergravity that are, in principle, parameterized by arbitrary functions of at least one variable.\footnote{For our explicit example solutions we will restrict attention to a sub-class of solutions parameterized by one integer, however by the above discussion, the broader family of these solutions is in principle described by arbitrary functions of at least one variable.} There is a long history of constructing smooth solutions in this theory \cite{Bena:2005va,Berglund:2005vb,Bena:2007kg}. However, while these solutions have non-trivial topology, they also have much symmetry; the solution spaces depend on several continuous parameters that describe the location of the topological bubbles. Until now it was not known how to construct smooth solutions in these theories parameterized by arbitrary continuous functions---such solutions were believed to exist only in supergravity theories in space-time dimensions greater than or equal to six, such as those in~\cite{Lunin:2001jy,Lunin:2002iz,Bena:2015bea,Bena:2016agb,Bena:2016ypk}.  Our map thus, in principle, yields the largest family, to date, of smooth solutions of five-dimensional ungauged supergravity. It also establishes that five-dimensional supergravity can capture smooth, horizonless solutions with black-hole charges, to a much greater extent than previously thought. 

The structure of the presentation is as follows. In Section 2 we introduce the class of six-dimensional BPS D1-D5-P geometries of interest, and the BPS equations that they satisfy. We work with asymptotically AdS$_3\times \IS^3$ geometries that can be written as a torus fibration, with the fiber coordinates $(v,\psi)$ asymptotically identified with (roughly speaking) the AdS$_3$ angular coordinate and the Hopf fiber coordinate of $\IS^3$ respectively.  We introduce maps that involve an $SL(2,\IQ)$, action on the torus fiber and a redefinition of the periodicities of these coordinates, and call these maps ``spectral transformations''.  
In Section~3 we illustrate the action of a particular spectral transformation on the example of the round, $\kappa$-wound multi-wound supertube solution. This transformation introduces KK monopole charge into the D1-D5 system.  We then recall the known U-duality that relates the D1-D5-KKM system to the MSW system on $\IT^6$ (or $\IT^2\times $K3).
In Section 4 we derive the effect of general $SL(2,\IQ)$ transformations on the six-dimensional metric and gauge fields. In Section~5 we apply our particular transformation to the D1-D5-P superstrata of~\cite{Shigemori:2013lta,Bena:2015bea,Bena:2016ypk}, and we work out an explicit example in detail in Section~6.
In Section~7 we investigate the question of whether there is a weakly coupled symmetric orbifold CFT in the moduli space of the MSW system, as there is for D1-D5.  When the compactification manifold, $\cM$, is $\IT^4$, the energetics of U(1) charged excitations can be inferred from a supergravity analysis~\cite{Larsen:1999dh}, and places strong constraints on the CFT, leading to a no-go theorem.  In Section 8 we discuss our results, and the appendices contain various technical details.

\section{D1-D5-P  BPS solutions and spectral transformations } 
\label{sec:D1-D5-P}

\subsection{D1-D5-P BPS solutions}
\label{sec:D1D5sols}

In the D1-D5-P frame, we work in type IIB string theory on $\cM^{4,1}\times \bbS^1\times \cM$, where $\cM$ is either $\IT^4$ or K3. We shall take the size of $\cM$ to be microscopic, and the $\bbS^1$ to be macroscopic. The $\bbS^1$ is parameterized by the coordinate $y$ which we take to have radius $R_y$,
\begin{equation}
y ~\sim~ y \,+\, 2 \pi R_y  \,. 
\label{yperiod}
\end{equation}
We reduce on $\cM$ and work in the low-energy supergravity limit. That is, we work with six-dimensional, $\Neql1$ supergravity coupled to two (anti-self-dual) tensor multiplets.  This theory contains all the fields expected from D1-D5-P string world-sheet calculations \cite{Giusto:2011fy}. The system of equations describing all \nBPS{8}, D1-D5-P solutions of this theory was found in  \cite{Giusto:2013rxa}; it is a generalization of the system discussed in \cite{Gutowski:2003rg,Cariglia:2004kk} and greatly simplified in \cite{Bena:2011dd}.
For supersymmetric solutions, the metric on $\cM$ takes the local form:
\begin{equation}
ds_6^2 ~=~    -\frac{2}{\sqrt{\cP}} \, (dv+\beta) \big(du +  \omega + \tfrac{1}{2}\, \mathcal{F} \, (dv+\beta)\big) 
~+~  \sqrt{\cP} \, ds_4^2(\cB)\,. \label{sixmet}
\end{equation}
The supersymmetry requires that all fields are independent of the null coordinate, $u$.    

In parameterizing D1-D5-P solutions it is standard to relate $u$ and $v$ to the circle coordinate, $y$, and a time coordinate $t$ via  
\begin{equation}
  u ~=~  \coeff{1}{\sqrt{2}} (t-y)\,, \qquad v ~=~  \coeff{1}{\sqrt{2}}(t+y) \,. \label{tyuv-1}
\end{equation}
However, there is some freedom in choosing such a relation, since the form of the metric (and the ansatz in general) is invariant under the shift
\begin{equation}
u' ~\equiv~ u ~-~ \coeff{1}{2}c_0 v\,,  \qquad  \cF' ~\equiv~  \cF ~+~ c_0 \,, \qquad  \omega'  ~\equiv~ \omega ~-~ \coeff{1}{2}\, c_0 \, \beta  \,.
\label{shiftinv}
\end{equation}
Using this freedom, we will shortly choose a different relation between $u$, $v$, $t$ and $y$ that is more natural for spectral transformations and for reduction to five dimensions.

While all the ansatz quantities may in principle depend upon $v$, throughout this paper we shall require the metric, $ds_4^2(\cB)$, on the four-dimensional spatial base, and the fibration vector, $\beta$, to be independent of $v$.  This greatly simplifies the BPS equations and, in particular, requires that the base metric be hyper-K\"ahler and that $d\beta$ be self-dual on $\cB$.  

The metric and tensor gauge fields are determined as follows. We introduce an index $I=1,\dots,4$, and an index $a$ that excludes $I=3$ (which plays a preferred role): $a=1,2,4$. The ansatz then contains four functions $Z_a$ and $\cF$, and four self-dual $2$-forms, $\Theta^{(I)}$, $I=1,\dots,4$. These can depend both upon the base, $\cB$, {\it and} upon the $v$ fiber.   The function, $\cF$, appears directly in (\ref{sixmet}) and the  warp factor, $\cP$, in the metric is given by 
\begin{equation}
\cP ~=~   Z_1\;\! Z_2 \,-\, Z_4^2\,. \label{Pform0}
\end{equation}
The vector field, $\beta$, defines  $\Theta^{(3)}$:
\begin{equation}
\Theta^{(3)} \equiv d \beta \,, \qquad  \Theta^{(3)}~=~ *_4 \Theta^{(3)} \,. \label{Theta3SD}
\end{equation}
The individual functions, $Z_a$, and the remaining $2$-forms, $\Theta^{(a)}$, encode the electric and magnetic components of the tensor gauge fields.  Recall that the $\Neql1$ supergravity multiplet contains a self-dual tensor gauge field, so that adding two anti-self-dual tensor multiplets means that the theory contains three tensor gauge fields. 

Roughly speaking, the pairs $(Z_1,\Theta^{(2)})$ and $(Z_2,\Theta^{(1)})$ describe the fields sourced by the D1 and D5 brane distributions.  The function, $\cF$, and the vector field, $\beta$, encode the details of the third momentum charge.  In the IIB description, the addition of $(Z_4,\Theta^{(4)})$ allows for a non-trivial NS-NS B-field as well as a linear combination of the R-R axion and four-form potential with all legs in the internal space $\cM$; these fields arise in D1-D5-P string world-sheet calculations \cite{Giusto:2011fy}, so are expected to be generically present.  For more details, see  \cite{Giusto:2013rxa}.  

The remaining simplified BPS equations come in two layers of linear equations. To write them, we denote by $d_{(4)}$ the exterior derivative on the four-dimensional base, and we define the operator, $D$, acting on a $p$-form with legs on the four-dimensional base (and possibly depending on $v$), by:
\begin{equation} 
D \Phi ~\equiv~ d_{(4)} \Phi ~-~ \beta \wedge {\partial_v{\Phi}} \,.  \label{Ddefn}
\end{equation} 
The first layer of equations determines the Maxwell data. 
For notational convenience throughout the paper, we work with a form of the BPS equations that is not explicitly covariant in the indices $a=1,2,4$. In particular, we will always label $\Theta_{4}$ with a downstairs index, while continuing to refer to these quantities collectively as $\Theta^{(a)}$; hopefully this will not cause confusion. The covariant form of the BPS equations is given in Appendix~\ref{app:covariant}.\footnote{To pass to the covariant form, one rescales $(Z_4,\Theta_4,G_4) \to (Z_4,\Theta_4,G_4)/\sqrt{2}$; more details are given in Appendix~\ref{app:covariant}.\label{foot:cov}} 
In our conventions, the first layer of the BPS equations takes the form
\begin{equation}
 \begin{aligned}
 *_4 D\dot{Z}_1 ~=~ &  D\Theta^{(2)}\,,\qquad D*_4 DZ_1  ~=~  -\Theta^{(2)} \! \wedge d\beta\,,
\qquad \Theta^{(2)} ~=~ *_4 \Theta^{(2)}\,, 
\cr
 *_4 D\dot{Z}_2  ~=~   &D\Theta^{(1)}\,,\qquad D*_4DZ_2  ~=~  -\Theta^{(1)} \! \wedge d\beta\,,
\qquad \Theta^{(1)} ~=~  *_4 \Theta^{(1)}\,, \\
 *_4 D \dot{Z}_4 ~=~ & D  \Theta_{4}\,,\qquad ~~D *_4 D Z_4 ~=~ -\Theta_{4}\wedge d\beta\,,
\qquad~~~ \Theta_{4} ~=~ *_4 \Theta_{4}\,.
\end{aligned}
\label{eqZTheta}
\end{equation} 
The second layer of equations determines the other parts of the metric in terms of the Maxwell data:
\begin{equation}
 \begin{aligned}
D \omega + *_4 D\omega + \mathcal{F} \,d\beta 
~=~ & Z_1 \Theta^{(1)}+ Z_2 \Theta^{(2)} -2\,Z_4 \thetafour \,,  \\ 
 *_4D*_4\!\Bigl(\dot{\omega} -\coeff{1}{2}\,D\mathcal{F}\Bigr) 
~=~& \partial_v^2 (Z_1 Z_2 - {Z}_4^2)  -(\dot{Z}_1\dot{Z}_2  -(\dot{Z}_4)^2 )
-\coeff{1}{2} *_4\! \big(\Theta^{(1)}\wedge \Theta^{(2)} - \thetafour\wedge \thetafour \big)\,.
\end{aligned}
\label{eqFomega}
\end{equation} 

Throughout this paper we shall take the base metric to be a Gibbons--Hawking (GH) metric:
\begin{equation}
ds_4^2 ~=~   V^{-1} \, (d \psi + A)^2   + V^{-1} ds_3^2 \,,  \qquad ~~  \nabla_{\!3}^2 V ~=~ 0\,, \qquad  \ast_3 dV ~=~ dA 
\label{GHmet2}
\end{equation}
where $ds_3^2$ is flat $\IR^3$, $\nabla_{\!3}^2$ is the $\IR^3$ Laplacian, and $\ast_3$ is the $\IR^3$ Hodge dual.
This leads to the following simple parametrization of solutions to (\ref{Theta3SD}): 
\begin{equation}
\beta ~=~
\frac{K^3}{V}\, (d \psi + A)~+~ \xi  
\,, \qquad ~~  \nabla_{\!3}^2 K^3 ~=~ 0\,, \qquad  \ast_3 dK^3 ~=~ - d\xi \,.
\label{betaform1}
\end{equation}

The choice of a GH base also leads to a convenient  decomposition of $\omega$ into its components parallel and perpendicular to the $\psi$-fiber:
\begin{equation}
\omega ~=~   \mu\,  (d \psi +A) ~+~ \varpi \,. \label{omform1}
\end{equation}
We record here the ansatz for the three-form field strengths in terms of the above data. A discussion of how these field strengths appear in the corresponding Type IIB ansatz may be found in  \cite{Giusto:2013rxa} and a simplified version without $(Z_4,\Theta_4)$ may be found in \cite{Bena:2011dd}. The BPS ansatz for the fluxes, where $ds_4^2$ and $\beta$ are $v$-independent, is given by:\footnote{Note that, following \cite{Niehoff:2013kia}, we have rescaled $\Theta^{(1,2)} \to \frac{1}{2} \Theta^{(1,2)}$ relative to the conventions of \cite{Bena:2011dd}. See also Footnote \ref{foot:cov}.} 
\begin{eqnarray}  
G^{(1)}  &=&  d \left[ - \frac{1}{2}\,\frac{Z_2}{\cal P}\,(du + \omega ) \wedge (dv + \beta)\, \right] ~+~\coeff{1}{2} *_4 D Z_2  
~+~ \coeff{1}{2}\,  (dv+ \beta) \wedge \Theta^{(1)} \,,    \cr
G^{(2)} &=&   d \left[ - \frac{1}{2}\,\frac{Z_1}{\cal P}\,(du + \omega ) \wedge (dv + \beta)\,   \right] ~+~  \coeff{1}{2} *_4 D Z_1  
~+~   \coeff{1}{2}\,  (dv+ \beta) \wedge \Theta^{(2)}  \,, \label{niceGform} \\
G_4 &=&   d \left[ - \frac{1}{2}\,\frac{Z_4}{\cal P}\,(du + \omega ) \wedge (dv + \beta)\,   \right] ~+~  \coeff{1}{2} *_4 D Z_4  
~+~   \coeff{1}{2}\,  (dv+ \beta) \wedge \Theta_4  \,.
\nonumber
\end{eqnarray}
These fields satisfy a twisted self-duality condition; since this is most conveniently expressed in covariant form \cite{Romans:1986er} (see also \cite{Bena:2016dbw}), we give it in Appendix \ref{app:covariant}.

\subsection{Canonical  transformations}
\label{sec:Coords}

As noted above around Eq.\;\eq{shiftinv}, there is some freedom in relating the $(u,v)$ coordinates to the time and spatial coordinates, $(t,y)$. As will shortly become clear, it will be convenient for us to use the following relation throughout this paper:
\begin{equation}
u ~=~ t \,,  \qquad v  ~=~  t+y \,; \qquad y ~\cong~  y +  2 \pi R_y\,.
\label{tyuv-2}
\end{equation}
One should note that the coordinates $(t,y)$ are the same as those in Eq.\;(\ref{tyuv-1}). To get from Eq.\;(\ref{tyuv-1}) to the above relation, one can make a shift (\ref{shiftinv}) with $c_0 =-2$, followed by a rescaling of $u'$ by $\frac{1}{\sqrt{2}}$ and $v$ by $\sqrt{2}$, together with accompanying rescalings of the ansatz quantities, as follows.
First, we take $c_0=-2$ in \eq{shiftinv} and define:
\begin{equation}
u' ~\equiv~ u + v\,,  \qquad \cF' ~\equiv~  \cF -2 \,, \qquad \omega' ~\equiv~ \omega +\beta  \,.
\label{shiftinv-app}
\end{equation}
Then we perform the rescalings:
\begin{equation}
\tilde{u} = \frac{u'}{\sqrt{2}} \,, \quad~ \tilde{v} =  \sqrt{2} \, v' \,, \quad~
\medtilde{\cF} = \frac{\cF'}{2} \,, \quad~  \medtilde{\omega} =\frac{1}{\sqrt{2}}  \;\! \omega'  \,, 
 \quad~ \medtilde{\beta} = \sqrt{2}\;\! {\beta}' \,,  
\quad~ \medtilde\Theta^{(a)} = \frac{\Theta^{(a)}}{\sqrt{2}}  \quad (a = 1,2,4) \,.
\label{rescalings}
\end{equation}
One then arrives at (\ref{tyuv-2}) by dropping all the tildes. 

Most importantly, with these re-scalings, the  ansatz for the metric,  the ansatz for the fluxes and the BPS equations remain unchanged: the factors of $\sqrt{2}$  cancel throughout.  Thus we are free to use either coordinate representation,  (\ref{tyuv-1}) or  (\ref{tyuv-2}), solve the BPS equations and substitute into the ans\"atze: both will produce BPS solutions.  The resulting solutions will of course be related by (\ref{shiftinv-app}) and (\ref{rescalings}).  We  illustrate this point with a simple supertube solution in Appendix~\ref{app:supertube}.

The $u=t$ parameterization is much more convenient when comparing six-dimensional solutions to five-dimensional solutions. Assuming that $\cal F$ is everywhere negative, as it will be in our solutions, one can write the metric as:
\begin{equation}
ds_6^2 ~=  \frac{1}{\sqrt{\cP}\, \mathcal{F}} \,(du +  \omega)^2 ~-~ \frac{\mathcal{F}}{\sqrt{\cP}}\, \big(dv+\beta + \mathcal{F}^{-1}  (du +  \omega)  \big)^2 
~+~  \sqrt{\cP} \, ds_4^2(\cB)\,. \label{sixmet-sq} 
\end{equation}
Since $\cal F$ is everywhere negative, the $v$ coordinate is everywhere spacelike. Solutions with an isometry along $v$ can be reduced on $v$ to obtain five-dimensional solutions. In this reduction, $u$ is the natural time coordinate in five dimensions. This is the advantage of the $u=t$ parameterization for our purposes.

Finally, to be clear: In this paper we will use (\ref{tyuv-2})  and $u=t$  will be kept fixed in all spectral transformations.

\subsection{General Spectral Transformations}
\label{SpecTrans}

Note that for solutions with a GH base, the six-dimensional solution in the form \eq{sixmet-sq}, \eq{GHmet2} is written as a double circle fibration, defined by $(v,\psi)$, over the $\IR^3$ base of the GH metric.
In this paper we will exploit a set of maps that involve coordinate transformations of the $(v,\psi)$ coordinates. We consider maps that act on $(v,\psi)$ with elements of SL$(2,\mathds{Q})$ and not just SL$(2,\mathds{Z})$, and so in general one must be careful to specify how the map acts on the lattice of periodic identifications of these coordinates. Our maps consist of a composition of a coordinate transformation and an accompanying redefinition of the lattice of coordinate identifications. 

The coordinate transformation component of our map is an SL$(2,\mathds{Q})$ map that transforms a solution written in terms of $(v,\psi)$ coordinates to a solution written in terms of new coordinates $(\hat v,\hat \psi)$. We parameterize the SL$(2,\mathds{Q})$ action by rational numbers $\mathbf{a},\mathbf{b},\mathbf{c}$ and $\mathbf{d}$ subject to $\mathbf{ad} -\mathbf{bc} =1$, as follows:
\begin{equation}
\frac{v}{R} ~=~  \mathbf{a}\;\! \frac{\hat v}{R}~+~ \mathbf{b} \, \hat \psi \,, \qquad\quad  \psi ~=~   \mathbf{c} \;\!  \frac{\hat v}{R} ~+~ \mathbf{d} \,\psi  \,, \qquad\quad R \;=\; \frac{R_y}{2} \,. \label{modmap1}
\end{equation}
For later convenience, in the above we have introduced the shorthand $R$ for the ratio of the periodicities of the $v$ and $\psi$ coordinates, so that the linear transformation acts on circles with the same period of $4 \pi$. We emphasize again that the coordinate $u$ is held fixed.

The lattice redefinition component of our map is as follows. We consider starting configurations for which the lattice of identifications is\footnote{For ease of exposition, here we suppress possible additional identifications that involve $\psi$ and an angle in the three-dimensional base; we will be more precise when we discuss explicit examples later.}
\begin{equation}
v  ~\cong~  v +  2 \pi R_y  \,,  \qquad\quad  \psi ~\cong~ \psi + 4\pi \,.
\label{v-psi-periods}
\ee
We define the new lattice of identifications of the new solution to be 
\begin{equation}
\hat{v}  ~\cong~  \hat{v} +  2 \pi R_y  \,,  \qquad\quad \hat\psi ~\cong~ \hat\psi + 4\pi \,,
\label{vhat-psihat-periods}
\ee
that is, the new lattice is not the one that would follow from making the coordinate transformation \eq{modmap1} on the original lattice \eq{v-psi-periods}, but is redefined to be~\eq{vhat-psihat-periods}. 

The fact that the lattice is redefined means that when the parameters of the map are non-integer, the maps are in general not diffeomorphisms and can modify the presence or absence of orbifold singularities in the spacetime, as has been observed in fractional spectral flow transformations~\cite{Giusto:2012yz}. We illustrate the above procedure by reviewing the example of fractional spectral flow transformations of multi-wound circular D1-D5 supertubes in Appendix \ref{app:lattice-spectral}.
We will refer to these maps as ``spectral transformations''.

Having made the above transformation, one can recast the metric and tensor gauge fields back into their BPS form but in terms of the new coordinates, $(\hat v, \hat \psi)$.   For example, one substitutes the coordinate change \eq{modmap1} into the metric (\ref{sixmet-sq}) and (\ref{GHmet2}), and then rewrites the result as: 
\begin{equation}
ds_6^2 ~=  \frac{1}{\sqrt{\widehat \cP} \, \widehat{\mathcal{F}}}  \,(du +  \hat \omega)^2 ~-~ \frac{\widehat{\mathcal{F}}}{\sqrt{\widehat\cP}}\, \big(d \hat v+\hat \beta +\widehat{\mathcal{F}}^{-1}  (du +  \hat \omega)  \big)^2 
~+~  \sqrt{\widehat \cP} \,\big( \widehat V^{-1} \, (d \hat \psi + \hat A)^2   + \widehat V^{-1} ds_3^2  \big) \,.  \label{recastsixmet} 
\end{equation}
This rearrangement of the background metric and tensor gauge fields in terms of the new fibers defines new `hatted'  functions and differential forms in terms of the old functions and forms.  We will derive the explicit transformation rules for the individual ansatz quantities in Section~\ref{sec:GenSpecTrans}, and use these rules to transform the family of superstrata solutions that we consider.

After this transformation the local metric is still the same as the original one, so the background is still \textit{locally} supersymmetric, and so the hatted ansatz quantities solve the BPS equations in the form~\eq{eqZTheta}, \eq{eqFomega}. More specifically, if the original functions and forms in the solution depend upon $(v,\psi)$ then that dependence must, of course, be transformed to $(\hat v, \hat \psi)$ using \eq{modmap1}, and the  BPS equations satisfied by the hatted quantities will be those of (\ref{eqZTheta}) and (\ref{eqFomega}) but with   $(v,\psi)$ replaced by $(\hat v, \hat \psi)$.  

While the transformed solution is still locally supersymmetric, it is possible that the redefined lattice of identifications may break some supersymmetry; indeed, we shall see that the transformation that we employ in the current work will break half of the eight real supersymmetries preserved by the D1-D5 circular supertube solution. 

\subsection{Five-dimensional solutions and spectral transformations}
 \label{sec:reduction-to-5D}

Solutions that are independent of $v$ can be dimensionally reduced from six to five dimensions. The BPS equations become those of $\Neql2$ supergravity coupled to three vector multiplets.  In particular, the description of the four vector fields of this theory involves  totally symmetric structure constants, $C_{IJK}$. Indeed, for the system we are considering one has
\begin{equation}
C_{123}  ~=~  1 \,, \qquad  C_{344}  ~=~  -2 \,,
\label{RecC}
\end{equation}
with all other independent components equal to zero.\footnote{One can convert this to the canonical normalization (in which $C_{344}=-1$) by the procedure described in Footnote \ref{foot:cov}.}

The complete family of smooth solutions that are also $\psi$-independent may then be written as follows \cite{Bena:2007kg}:\footnote{Note that our convention for $M$ differs from that of~\cite{Bena:2007kg} by a factor of 2.} 
\begin{align}
\Theta^{(I)}  &~=~ dB^I \,, \quad~~~ B^I ~\equiv~  \frac{K^I}{V}\, (d \psi + A) \,+\, \xi^I  \,,
\quad~~ \nabla_{\!3}^2 K^I ~=~ 0\,, \quad~~ \ast_3 dK^I ~=~ - d\xi^I  \,; \cr
 \qquad Z_I  &~=~ L_I \,+\, \frac{1}{2}\, C_{IJK}  \frac{K^{J}K^{K}}{V}     \,, \qquad ~~  \nabla_{\!3}^2 L_I ~=~ 0 \,,
\label{harmfns} \\
\mu &~=~  \frac{M}{2} + \frac{K^I L_I}{2 \;\! V} + \frac{1}{6} \, C_{IJK}\,  \frac{K^I K^J K^K}{V^2}  \,, \qquad~ \nabla_{\!3}^2 M ~=~ 0\,,
\nonumber
\end{align}
with $\varpi$ determined by 
\begin{equation}
 \ast_3 d \varpi ~=~ \frac12 \left( V d M -
M d V +   K^I d L_I - L_I d K^I \right)\,.
\label{omegeqn}
\end{equation}
We now reduce the metric, (\ref{sixmet-sq}), on the $v$-fiber. Following  (\ref{tyuv-2}), we set $u=t$, and we relabel $\omega$ and $\cal F$ in terms of their more standard five-dimensional analogs: 
\begin{equation}
\mathbf{k}   ~\equiv~ \omega  \,, \qquad  Z_3 ~\equiv~ - \cal F \,.
\label{tkZ3defn}
\end{equation}
This yields the standard five-dimensional metric:
\begin{equation}
ds_5^2   ~=~  -(Z_3 \,\cP )^{-\frac{2}{3}} \, (dt +  \mathbf{k})^2   ~+~  (Z_3 \,\cP )^{\frac{1}{3}} \,  ds_4^2(\cB) \,.
\label{fivemet}
\end{equation}

\subsection{Spectral Transformations for $v$-independent solutions}
\label{sec:SFGT1}

The role of  $SL(2,\ZZ)$ spectral transformations on $(v,\psi)$ was studied in detail for $v$-independent solutions in 
\cite{Bena:2008wt}.    In particular, the  spectral transformations could be reduced to transformations on the  the harmonic functions $V, K^I, L_I$ and $M$.   Moreover, from the five-dimensional perspective, any of the Maxwell fields can be promoted to the Kaluza-Klein field of the six-dimensional formulation and so there as many different $SL(2,\ZZ)$  spectral transformations as there are vector fields.  Moreover, these $SL(2,\ZZ)$ actions do not commute and, in fact, generate some even larger sub-group of the U-duality group.  

The original study of spectral transformations was made for the system with two vector multiplets ($Z_4 \equiv 0$ and $\Theta_4 \equiv 0$) but the results can be recast in a form that is valid for five-dimensional $\Neql 2$ supergravity coupled to $(\nv-1)$ vector multiplets, and so we will give the relevant general results. 

The  spectral transformations considered in \cite{Bena:2008wt} included two important sub-classes: ``gauge transformations'' and ``generalized spectral flows''. 
A gauge transformation is generated by choosing one of the Maxwell fields as the KK field,  then leaving $\psi$ fixed and shifting $v$ by a multiple of $\psi$.   The choices of uplift lead to  $\nv$ gauge parameters, $g^I$,  and the gauge transformations reshuffle the harmonic functions according to:
\bea
\widehat{V} &=& V \,, \qquad \widehat{K}^I ~=~ K^I + g^I V \,, \cr
\widehat{L}_I &=&L_I - C_{IJK} g^J K^K -\frac12 C_{IJK} g^J g^K V \,, \cr
\medhat{M} &=& M- g^I L_I + \frac{1}{2} C_{IJK} g^I g^J K^K
+ \frac{1}{3!} C_{IJK} g^I g^J g^K  V \,.
\label{eq:gauge-harm}
\eea 
While this is a highly non-trivial action on the harmonic functions, this transformation leaves the physical fields, $Z_I, \Theta^{(I)}$, $\mu$ and $\varpi$ invariant, and hence their designation as gauge transformations. 

Spectral flows are induced by keeping  $v$ fixed and shifting $\psi$ by a multiple of $v$.  Again there are $\nv$ ways to do this with $\nv$ parameters, $\gamma_I$, resulting in the full family of generalized spectral flow transformations.  They act on the harmonic functions as follows:
\bea
\medhat{M} &=& M \,, \qquad \widehat{L}_I ~=~ L_I - \gamma_I M \,, \cr
\widehat{K}^I &=& K^I - C^{IJK} \gamma_J L_K + \frac12 C^{IJK} \gamma_J \gamma_K M \,, \cr
\widehat{V} &=& V + \gamma_I K^I - \frac{1}{2} C^{IJK} \gamma_I \gamma_J L_K
+ \frac{1}{3!} C^{IJK} \gamma_I \gamma_J \gamma_K  M \,.
\label{eq:SF-harm}
\eea
In contrast to the gauge transformations, these transformations have a complicated and very non-trivial action on the five-dimensional physical fields (see \cite{Bena:2008wt}).  

In our conventions, the polarization direction $I=3$ in Eq.\;\eq{eq:SF-harm} corresponds to the (Kaluza-Klein) vector field in five dimensions that lifts to metric in six dimensions. We reserve the term ``spectral flow'' for generalized spectral flows in this polarization direction.

Spectral flow transformations have the same effect as the following large coordinate transformation, where the new coordinates are denoted with a hat:
\bea
\psi &=& \hat\psi  + \gamma_3 \;\! \hat{v} \,, \qquad v ~=~ \hat{v} \,,
\eea
and where the other coordinates are invariant. In the D1-D5 system CFT, the world volume of the CFT lies along the $v$-fiber and translations along this fiber are generated by the Hamiltonian, $L_0$. The $\psi$-fiber lies transverse to the D1 and D5 branes and so represents an $\cR$-symmetry transformation. The above transformation is thus a CFT spectral flow; for a more detailed discussion, see~\cite{Giusto:2004id,Giusto:2004ip,Giusto:2012yz}.

Similarly, a gauge transformation in the polarization direction 3 with parameter $g^3$ has the same effect as the following coordinate transformation, where again the new coordinates are denoted with a hat:
\bea
v &=& \hat{v} + g^3 \;\! \hat{\psi} \,,
\qquad \psi ~=~ \hat{\psi} \,,
\eea
and where the other coordinates are invariant. Note that, when the world-volume of the CFT lies along the $v$-fiber, this seemingly trivial (from a supergravity point of view) transformation does not appear to have a simple interpretation in the dual CFT.  The transformation would appear to reorient the world-volume of the CFT, and the question of whether there is any sensible holographic interpretation of this gravity transformation remains somewhat mysterious. 

However, if one interchanges the roles of $\psi$ and $v$ such that the world-volume of the CFT lies along the $\psi$-fiber, and the Hopf fiber of the S$^3$ lies along $v$, then (in our conventions) the above gauge transformation would correspond to spectral flow in the left-moving sector of the CFT.

\newpage
\section{The multi-wound supertube and mapping D1-D5 to MSW}
\label{sec:supertube}
 
Before proceeding to general spectral transformations, it is very instructive to see how spectral transformations act on one of the most important $v$-independent BPS solutions: the  multi-wound  supertube~\cite{Balasubramanian:2000rt,Maldacena:2000dr,Mateos:2001qs,Emparan:2001ux}.  We start with its standard formulation as the smooth geometry of the  D1-D5 supertube, and we map it to M-theory with an SL$(2,\QQ)$ spectral transformation and a U-duality transformation. The SL$(2,\QQ)$ spectral transformation introduces a KKM charge along the Hopf fiber of the S$^3$, and the D1, D5 and KKM charges transform under the U-duality into three independent M5-brane charges underlying an MSW string, where the resulting configuration is a particular form of that string, with specific dissolved M2-brane charges and specific angular momenta.

There is the following interesting conundrum:  the D1-D5 supertube is \nBPS{4}, preserving eight supersymmetries, while the MSW string is \nBPS{8} and preserves only four supersymmetries. We reconcile this difference by carefully examining the lattice of identifications, and showing that our transformation accounts for the change in the number of supersymmetries.

\subsection{The multi-wound D1-D5 supertube configuration} 
\label{ss:background}

The canonical starting point for the multi-wound, circular D1-D5 supertube is the ansatz \eq{sixmet} with a time coordinate, $t$, and the asymptotic $\bbS^1$ coordinate, $y$, related to the coordinates $u,v$ via (\ref{tyuv-1}) and with $\mathcal{F} = 0$.  However, as we stipulated earlier, we are going to  use (\ref{tyuv-2}), and the transformations (\ref{shiftinv-app}) and (\ref{rescalings}) then imply that one must take  $\mathcal{F} = -1$.   The  precise relation between the supertube with  (\ref{tyuv-1}) and the supertube with  (\ref{tyuv-2}) is given in Appendix \ref{app:supertube}. 

The $\kappa$-wound supertube is a two-centered configuration defined by the following harmonic functions:
\bea
V &=& \frac{1}{r_+}  \,, \qquad K^1 ~=~ K^2 ~=~ 0 \,, \qquad 
K_3 ~=~ \frac{\kappa R_y}{2} \left( \frac{1}{r_-} \,-\,   \frac{1}{r_+}\right) \,, 
\cr
L_1 &=& \frac{Q_1}{4 \;\! r_-} \,, \qquad L_2 ~=~\frac{Q_5}{4 \;\! r_-} \,,\qquad L_3 ~=~1 \,, \qquad 
M ~=~ \frac{Q_1 Q_5}{8 \;\! \kappa R_y \;\! r_-} \,.
\label{eq:ST-harm-fns}
\eea
We write
\begin{equation}
\Sigma ~\equiv~ 4 r_- ~\equiv~ (r^2 + a^2 \cos^2 \theta) \,, \qquad   \Lambda ~\equiv~4 r_+ ~\equiv~(r^2 + a^2 \sin^2 \theta)  \,. \label{LamSigdefn}
\end{equation}
The base metric is flat $\mR^4$, which we write as
\begin{equation}
ds_4^2 ~=~ \Sigma\, \bigg( \frac{dr^2}{(r^2 + a^2)} ~+~ d \theta^2 \bigg)  ~+~ (r^2 + a^2) \sin^2 \theta \, d\varphi_1^2 ~+~ r^2  \cos^2 \theta \, d\varphi_2^2  \,.
\label{basemet}
\end{equation} 
The remaining ansatz quantities for the supertube are then: 
\bea
Z_1  &=& \frac{Q_1}{\Sigma} \,, \qquad Z_2  ~=~ \frac{Q_5}{\Sigma}\,, \qquad 
\cF ~=~ -1 \,, \qquad Z_4  ~=~0 \,;  \qquad\quad \Theta^{(I)} ~=~ 0\,, \quad I = 1,2,4  \,;
\cr
\beta &=& \frac{ \kappa R_y a^2 }{\Sigma} \;\! ( \sin^2 \theta \, d\varphi_1-   \cos^2 \theta \, d\varphi_2 )\,, \qquad  
\omega  ~=~\frac{\kappa R_y a^2 }{\Sigma} \sin^2 \theta \, d\varphi_1 \,. 
\label{eq:betaomegaST}
\eea
The parameters are subject to the following regularity condition:
\begin{equation}
Q_1 Q_5 ~=~  \kappa^2 R_y^2 \;\! a^2 \,. \label{streg}
\end{equation}
This solution may be written in Gibbons--Hawking form by defining new coordinates, $(\psi, \phi, \vartheta_{-})$ via:
\bea
~\sin \coeff{1}{2} \vartheta_{+} \;=\; \frac{(r^2 + a^2)^{1/2}}{\Lambda^{1/2}}\, \sin \theta \,, \quad~  \cos \coeff{1}{2}  \vartheta_{+} \;=\; \frac{r}{\Lambda^{1/2}}\, \cos \theta  \,, \label{coords1} 
\quad~ 
\psi \;=\; \varphi_1 +\varphi_2 \,, \quad~ \phi \;=\; \varphi_2 -\varphi_1 \,.
\nonumber
\eea
One then has 
\begin{eqnarray}
ds_4^2 &=&  V^{-1} \, (d \psi + A)^2   + 
V^{-1} \Big[ dr_+^2 \,+\, r_+^2 \, (d \vartheta_{+}^2 \,+\, \sin^2 \vartheta_{+} \;\! d\phi^2)  \Big]\,,  \label{GHmet2-a} \\
V &=&\frac{1}{r_+} ~=~  \frac{4}{\Lambda}  \,, \qquad A~=~ \cos \vartheta_{+} \,d \phi ~=~ \frac{\big((r^2 +a^2) \sin^2 \theta  -  r^2 \cos^2 \theta \big)}{\Lambda}   \,(d\varphi_1 -d\varphi_2) \,. 
\nonumber
\end{eqnarray}
The decomposition of $\beta$  in  (\ref{betaform1}) is given by $K^3$ in \eq{eq:ST-harm-fns} and
\begin{align}
\xi ~=~   \frac{\kappa  R_y a^2}{\Lambda \, \Sigma}\,(2\;\! r^2+a^2)\, \sin^2 \theta  \cos^2 \theta \,(d\varphi_1 -d\varphi_2) \,. \label{Kxiform1}
\end{align}
The quantities $r_\pm$ measure the distances in the flat three-dimensional base between two centers, defined by $r_\pm=0$; one can choose Cartesian coordinates in which we have 
\begin{equation}
r_\pm ~=~  \sqrt{y_1^2 + y_2^2 + (y_3 \mp c)^2} \,, \qquad c = \coeff{1}{8} \, a^2 \,. \label{cartform1}
\end{equation}
This solution describes a $\kappa$-wound supertube, whose KKM dipole 
moment is $\kappa$. Given the above choice of gauge for the one-form $A$, the lattice of identifications for this solution is generated by\footnote{By shifting $\psi$ to $\psi' = \psi \pm \phi$ one obtains $A=(\cos \vartheta_{-}\mp 1)d \phi$, as appropriate for smooth coordinate patches around the North and South Poles of the $\IS^2$ respectively. The identifications on $\psi'$ and $\phi$ are then simply $\psi' \cong \psi' + 4\pi$ at fixed $\phi$, and $\phi \cong \phi+2\pi$ at fixed $\psi'$.
\label{foot:psi-phi-lattice}}
\begin{eqnarray}
\left(y,\psi,\phi \right) &\sim& \left(y+2\pi R_y ,\,\psi,\,\phi \right) , \cr
\left( y,\psi,\phi  \right) &\sim& \left(y,\,\psi+4\pi,\,\phi \right) , 
\label{eq:lattice-D1D5} \\
\left( y,\psi,\phi  \right) &\sim& \left( y,\,\psi+2\pi, \, \phi+2\pi \right) . \nonumber
\end{eqnarray}
There is a $\ZZ_\kappa$ orbifold singularity at the supertube locus, as we will now review.

Introducing the coordinates
\begin{eqnarray}
r \!\!&=&\!\! a \sinh \zeta \,, \qquad \eta \,=\, 2 \theta \,, \qquad \tilde{y} \,=\, \frac{y}{\kappa R_y} \,, \qquad  \tilde{t} \,=\, \frac{t}{\kappa R_y} \,, 
\label{eq:tildecoords}
\end{eqnarray}
the metric can be written as
\begin{eqnarray}
ds_6^2  &=&  \sqrt{Q_1 Q_5}\left[ -  \cosh^2 \zeta \, {d\tilde{t}}^2 + d \zeta^2
 +   \sinh^2 \zeta \,  d\tilde{y}^2 \right] 
\label{AdStimesS-2b}\\
&&
{}\hspace{-6mm} + \frac{\sqrt{Q_1 Q_5}}{4} \left[
\Big( \big[d\psi - (d\tilde{t}+d\tilde{y}) \big] + \cos \eta \;\! \big[ d\phi +(d\tilde{t}-d\tilde{y}) \big] \Big)^2  
+  d\eta^2+ \sin^2 \eta \;\! \big[ d\phi +(d\tilde{t}-d\tilde{y}) \big]^2  
  \right] .
\nonumber
\end{eqnarray}
Under the further change of coordinates
\be
\tilde{\psi} \,=\, \psi - (\tilde{t}+\tilde{y}) \,, \qquad\quad  \tilde{\phi} \,=\, \phi + (\tilde{t}-\tilde{y})\,,
\label{eq:tildecoords2}
\ee
we observe that the metric is locally AdS$_3\times$S$^3$,
\begin{equation}
ds_6^2  ~=~  \sqrt{Q_1 Q_5}\left[ -  \cosh^2 \zeta \, {d\tilde{t}}^2 + d \zeta^2
 +   \sinh^2 \zeta \,  d\tilde{y}^2 + \frac{1}{4}\left(
\left(d \tilde\psi  + \cos \eta \, d \tilde\phi \right)^2  
+  d\eta^2+ \sin^2 \eta \, d \tilde\phi^2  
\right)
  \right].
\label{AdStimesS-2c}
\end{equation}
The lattice of identifications in terms of the local AdS$_3\times$S$^3$ coordinates is generated by
\begin{eqnarray}
\left( \tilde{y},\tilde{\psi},\tilde{\phi} \right) &\sim& \left( \tilde{y}+\frac{2\pi}{\kappa},\,\tilde{\psi}-\frac{2\pi}{\kappa},\,
\tilde{\phi}-\frac{2\pi}{\kappa}\right) , \cr
\left( \tilde{y},\tilde{\psi},\tilde{\phi} \right) &\sim& \left( \tilde{y},\,\tilde{\psi}+4\pi,\,\tilde{\phi} \right) , 
\label{eq:lattice-D1D5-2} \\
\left( \tilde{y},\tilde{\psi},\tilde{\phi} \right) &\sim& \left( \tilde{y},\,\tilde{\psi}+2\pi, \, \tilde{\phi}+2\pi \right). \nonumber
\end{eqnarray}
When $\kappa=1$, the full geometry is simply global AdS$_3\times$S$^3$, and when $\kappa>1$, the first identification above is an orbifold identification that combines the AdS$_3$ and the S$^3$, and that gives rise to a $\mathds{Z}_\kappa$ orbifold singularity at the location of the supertube, ($r=0,\theta=\pi/2$)~\cite{Balasubramanian:2000rt,Maldacena:2000dr} (see also \cite{Giusto:2012yz}).

\subsection{Spectral transformations of the supertube}
\label{ss:ST_SpecTrans}

We now perform an SL$(2,\QQ)$ map on the above multi-wound supertube configuration, that maps the solution to a form in which it can be straightforwardly dualized to the MSW frame. Our $SL(2,\QQ)$ map can be decomposed into a product of gauge and spectral flow transformations, and it will be instructive to go through these steps.

We first perform a gauge transformation with parameters $(g^1,g^2,g^3) = (0,0,\frac12 \kappa R_y)$. The resulting harmonic functions are:
\bea
V &=& \frac{1}{r_+} \,, \qquad K^1 ~=~ K^2 ~=~ 0 \,, \qquad K^3 ~=~  \frac{\kappa R_y}{2}\frac{1}{r_-}  \,, \cr
L_1 &=& \frac{Q_1}{4 \;\! r_-} \,, \qquad L_2 ~=~\frac{Q_5}{4 \;\! r_-} \,, \qquad 
L_3 ~=~1 \,, \qquad M ~=~ - \frac{\kappa R_y}{2} + \frac{Q_1 Q_5}{8 \;\! \kappa  R_y \;\! r_-} \,.
\eea
We next perform a (fractional) spectral flow transformation with parameters $(\gamma_1,\gamma_2,\gamma_3) = (0,0,-2/(\kappa R_y))$.
The resulting harmonic functions are:
\bea
V &=& \frac{1}{r_+} - \frac{1}{r_-} \,, \qquad K^1 ~=~  \frac{Q_5}{2 \;\! \kappa R_y} \frac{1}{r_-}  \,, 
\qquad K^2 ~=~  \frac{Q_1}{2 \;\! \kappa R_y}\frac{1}{r_-}  \,, \qquad K^3 ~=~  \frac{\kappa R_y}{2}\frac{1}{r_-}   \,, 
\cr
L_1 &=& \frac{Q_1}{4 \;\! r_-} \,, \qquad L_2 ~=~\frac{Q_5}{4 \;\! r_-}  \,, \qquad 
L_3 ~=~\frac{Q_1 Q_5}{4 \;\! \kappa^2 R_y^2 \;\! r_-} \,, \qquad M ~=~ - \frac{\kappa R_y}{2} + \frac{Q_1 Q_5}{8 \;\! \kappa  R_y \;\! r_-} \,. \quad 
\label{eq:sf-gauge}
\eea
Finally, we perform a gauge transformation with parameters $(g^1,g^2,g^3) = (\frac{Q_5}{4\kappa R_y},\frac{Q_1}{4\kappa R_y},\frac{\kappa R_y}{4})$. The resulting harmonic functions are (here $I=1,2,3$ and we employ notation mod 3 for the $I$ indices):
\begin{eqnarray}
 V &=&  \frac{1}{r_+} - \frac{1}{r_-} \,, \qquad\qquad\qquad\qquad~\,   
K^I ~=~ \frac{k^I}{2} \left( \frac{1}{r_+} +  \frac{1}{r_-} \right) \,,  \cr
 L_I &=& - \frac{k^{I+1} k^{I+2}}{4}  \left( \frac{1}{r_+} - \frac{1}{r_-} \right) \,,
 \qquad 
M ~=~  \frac{k^1k^2k^3}{8} \left( \frac{1}{r_+} +  \frac{1}{r_-} \right)  - \frac{k^1k^2k^3}{2c} \,.
\label{eq:functdec}
\end{eqnarray}
where we have defined
\bea
k^1 ~\equiv~ \frac{Q_5}{2 \;\! \kappa R_y} \,, \qquad k^2 ~\equiv~ \frac{Q_1}{2 \;\! \kappa R_y} \,, \qquad k^3 ~\equiv~ \frac{\kappa  R_y}{2} \,.
\label{eq:kItoQ15k}
\eea
These harmonic functions are those that describe the MSW maximally-charged Ramond ground state solution in five dimensions (related by right-moving spectral flow to the NS vacuum)~\cite{Denef:2007yt,deBoer:2008fk}, as reviewed in Appendix \ref{app:MSW-vac}. We will review the duality map momentarily.

The combination of these transformations corresponds to the following SL$(2,\QQ)$ map on the coordinates:
\begin{equation}
  \frac{v}{\kappa R_y}  ~=~ \frac{1}{2}\, \hat \psi  \,, \qquad  
	\psi = \frac{1}{2 }\,\hat \psi -   \frac{2 }{\kappa R_y}\hat v \,,  \qquad\quad (u,\phi) ~\, \mathrm{fixed}
		\,, \label{MSW2map}
\end{equation}
and where as discussed above, we redefine the new lattice of identifications. The new lattice is generated by the appropriate smooth identifications in the M-theory frame; combining \eq{vhat-psihat-periods} with the appropriate smooth identification on $\phi$ as discussed around \eq{eq:psi-phi-lattice-MSW-app}, the new identifications are:
\begin{equation}
\hat{v}  ~\cong~  \hat{v} +  2 \pi R_y  \,,  \qquad\quad \hat\psi ~\cong~ \hat\psi + 4\pi \,, \qquad\quad
\hat\phi ~\cong~ \hat\phi + 2\pi  \,,
\label{vhat-psihat-periods-2}
\ee
where for each of the three generators in this equation, one holds the other two periodic coordinates in the equation fixed. 

Note that in the tilded coordinates, defined in \eq{eq:tildecoords2}, the coordinate transformation becomes
\begin{equation}
  \frac{v}{\kappa R_y}  ~=~ \frac{\hat \psi }{2}\,  \,, \qquad  
 \tilde\psi \;= \; -   \frac{\hat v}{k^3} \,,    \qquad\quad  (u,\phi) ~\, \mathrm{fixed} \,,
	\label{MSW2map-2}
\end{equation}
which can be described as a (fractional) ``spectral interchange'' transformation~\cite{Niehoff:2013kia} between rotating versions of the AdS$_3$ circle coordinate and the Hopf fiber coordinate of the S$^3$. Indeed, under this transformation, the metric \eq{AdStimesS-2b} transforms to: 
\bea
ds_6^2 &=&  \sqrt{Q_1 Q_5}\left[ -  \cosh^2 \zeta \, {d\tilde{t}}^2 + d \zeta^2
+   \sinh^2 \zeta \;\! d\varphi^2
+ \frac{1}{4}\left(
\left(  \frac{d \hat v}{k^3}  + \cos \eta \, d \tilde\phi \right)^2  
+  d\eta^2+ \sin^2 \eta \, d \tilde\phi^2  
\right)
  \right]  \cr &&
	\label{AdStimesShat}
\eea
where we define
\be
\varphi ~\equiv~   \frac{\hat\psi}{2} -  \tilde{t}    \,.
\ee
It is interesting to re-interpret the coordinate transformation \eq{MSW2map-2} in terms of the D1-D5 and MSW CFTs.  The relation between $\tilde{\psi}$ and $\psi$ in \eq{eq:tildecoords2} corresponds to spectral flow in the left-moving sector of the D1-D5 CFT; the solution in terms of $\psi$ corresponds to a particular Ramond ground state, while the solution in terms of $\tilde{\psi}$ corresponds to the NS vacuum of the left-moving sector. (The analogous statement holds for $\tilde{\phi}$ and $\phi$ in terms of spectral flow in the right-moving sector of the CFT.)
The coordinate transformation in the tilded coordinates \eq{MSW2map-2} is interesting, as we see from it that $\hat{v}$ is a rescaled version of $\tilde{\psi}$, the left-moving NS sector coordinate.

One can paraphrase these observations by describing the  metric \eq{AdStimesShat} as being written in NS-NS sector coordinates, corresponding to the NS-NS vacuum of the dual CFT state, which has
\be
L_0 ~=~ \bar{L}_0 ~=~ 0 \,.
\ee
If one rewrites $\tilde\phi$ in terms of $\phi$, one can describe the metric as being expressed in NS-R sector coordinates, and corresponding to the NS-R ground state obtained from the NS-NS vacuum via right-moving spectral flow with parameter 1/2, which has 
\be
L_0 ~=~ 0 \,, \qquad \bar{L}_0 ~=~ \frac{c}{24} \,.
\ee
We will ultimately reduce on the $\hat{v}$ fiber, and these quantum numbers will correspond respectively to the NS vacuum and the maximally-charged R ground state of the MSW CFT (which we again emphasize are related by right-moving spectral flow).

Let us analyze the lattice of identifications \eq{vhat-psihat-periods-2} that has resulted from our transformation. The AdS$_3$ angle coordinate $\varphi$ has period $2\pi$, which is the correct periodicity for a smooth global AdS$_3$. The combination that appears in place of the Hopf fiber of the $\IS^3$ is $\hat{v}/k^3$, which has period $4\pi/\kappa$, corresponding to a smooth $\mathds{Z}_\kappa$ quotient of the Hopf fiber, appropriate for the decoupling limit of a D1-D5-KKM configuration with KKM charge $\kappa$.

Note now that the relation between the dimensionful parameters $Q_1$, $Q_5$ and the integer number of D1 and D5 branes $n_1$, $n_5$ that correspond to this solution is changed, relative to the solution without KKM charge. The Gaussian integral defining the charges is now done over a range of the Hopf fiber coordinate that is smaller by a factor of $\nk$, so that the actual new supergravity charges are 
\bea
\medtilde{Q}_1 = \frac{Q_1}{\nk} \,, \qquad \medtilde{Q}_5 = \frac{Q_5}{\nk} \,.
\label{eq:KKMrescaledcharges}
\eea
Recalling the usual relation between $\medtilde{Q}_{1,5}$ and $n_{1,5}$:
\bea
\medtilde{Q}_1 = \frac{g_s \alpha'{}^3 n_1}{V_4} \,, \qquad \medtilde{Q}_5 = g_s \alpha' n_5 \,,
\eea
we see that the relation between the integer brane numbers on the two sides of the map is 
\bea
n_1 = \frac{N_1}{\nk} \,, \qquad  n_5 = \frac{N_5}{\nk} \,.
\label{eq:KKMrescaledintegercharges}
\eea
We note that using \eq{eq:KKMrescaledcharges}, the relation \eq{eq:kItoQ15k} becomes
\bea
k^1=\frac{\medtilde{Q}_5}{2 R_y} \,, \qquad k^2 = \frac{\medtilde{Q}_1}{2 R_y} \,, \qquad k^3 = \frac{\kappa  R_y}{2} \,.
\label{eq:kItoQ15k-2}
\eea
so that, up to constant factors, the parameters $k^I$ correspond to the D1, D5 and KKM charges, which map to the three different M5 charges in the M5-M5-M5 duality frame.

The other effect of the redefined lattice of identifications is that, in the asymptotically AdS$_3\times\IS^3$ geometry, it breaks the $SU(2)_L \times SU(2)_R$ symmetry of the $\IS^3$  down to $U(1)_L \times SU(2)_R$.  Since this is the $\cR$-symmetry, it must break the $\Neql4$ superalgebra of the left-moving sector down to an  $\Neql2$ superalgebra with this $U(1)_L$ $\cR$-symmetry.  Since these remaining supercharges are charged under the translations along the Hopf fiber, they will not survive the dualization to M-theory and thus the effect of reassigning the lattice identifications and compactifying is to break all the left-moving supersymmetries even in the ground-state configuration we are studying here.

\subsection{Mapping from D1-D5-KKM to M-theory}
\label{ss:TypeIIB-to-Mtheory}

The duality map from D1-D5-KKM to M-theory involves T-duality on the Hopf fiber of the $\IS^3$ and two directions in the $\IT^4$, followed by an M-theory lift.\footnote{For early works on reduction and T-duality along Hopf fibers, see~\cite{Duff:1998us,Itzhaki:1998uz,Boonstra:1998yu}.} This results in a solution which asymptotically has a compact $\IT^6$. The effect of these dualities from the point of view of the lower-dimensional theory is encoded in a dimensional reduction on the Hopf fiber of the $\IS^3$, to five dimensions (see e.g.~\cite{Emparan:2006mm,Virmani:2012kw}). 

The SL$(2,\QQ)$ transformation we have performed means that the ansatz quantities have already been rearranged to make this step straightforward: the coordinate $\hat{v}$ is precisely the Hopf fiber of the $\IS^3$.

The metric \eq{AdStimesShat} can be written (using an obvious shorthand) as 
\bea
ds_6^2&=& \sqrt{ Q_1 Q_5} \, ds^2_{\rm{AdS_3}} 
+ \frac{\sqrt{Q_1 Q_5}}{4(k^3)^2} \left( d\hat{v} + k^3 \cos\eta \;\! d\tilde\phi\right)^2
+ \frac{\sqrt{Q_1 Q_5}}{4}  \, ds^2_{{\rm S}^2} \,.
\eea
The reduction ansatz for the six-dimensional metric takes the form 
\be
ds_6^2 ~\equiv~ e^{-3\mathcal{A}} (d\hat{v} + \hat{A}^{(3)})^2 +  e^{\mathcal{A}}\, ds_5^2 \,,
\ee
so we obtain
\bea
ds_5^2 ~=~ \left( \frac{Q_1 Q_5}{2 \;\! k^3} \right)^{2/3} ds^2_{\rm{AdS_3}}
 + \frac{1}{4}\left( \frac{Q_1 Q_5}{2 \;\! k^3}  \right)^{2/3}ds^2_{{\rm S}^2} \,.
\eea
Using the relation \eq{eq:kItoQ15k}, and as reviewed in Appendix~\ref{app:MSW-vac}, this becomes exactly 
the decoupled M5-M5-M5 metric in five dimensions that results from the set of harmonic functions~\eq{eq:functdec}~\cite{Denef:2007yt} (see also~\cite{deBoer:2008fk,Bena:2010gg,Bena:2013ora}):
\begin{equation}
ds_5^2 ~=~ R_1^2 \Big( - \cosh^2\zeta \,  d\tau^2 + d\zeta^2 +  \sinh^2 \zeta \, d\varphi^2 \Big) ~+~  R_2^2 \left(   d \eta ^2 + \sin^2\eta  \, d\tilde\phi^2 \right)  ,
 \label{AdS3S2}
 \end{equation}
with
\begin{equation}
  R_1~=~  2 R_2 ~=~ 4 (k^1k^2k^3)^{1/3} \,.
 \label{Radii}
 \end{equation}
Note that the smooth $\mathds{Z}_{\nk}$ quotient on the Hopf fiber has migrated into an M5 charge, and thus a parameter in the warp factors, and the five-dimensional  solution is smooth AdS$_3\times\IS^2$ with standard coordinate identifications.

\section{General SL$(2,\mathds{Q})$ transformations in six dimensions}
\label{sec:GenSpecTrans}

Having understood how to map the ground-state of the D1-D5 system onto that of the MSW string, we now wish to extend our results to   the transformation to families of left-moving excitations.    This includes   spectral transformations of superstrata  \cite{Bena:2015bea,Bena:2016agb,Bena:2016ypk}.   

We  start more generally by considering a generic BPS background that can depend on all of the coordinates, except, of course, $u$.  We recall our parameterization of the general spectral transformations on $(v,\psi)$ from Eq.\;\eq{modmap1}:
\begin{equation}
\frac{v}{R} ~=~  \mathbf{a}\;\! \frac{\hat v}{R}~+~ \mathbf{b} \, \hat \psi \,, \qquad\quad  \psi ~=~   \mathbf{c} \;\!  \frac{\hat v}{R} ~+~ \mathbf{d} \,\psi  \,, \qquad\quad R \;=\; \frac{R_y}{2} \,. \label{modmap2}
\end{equation}
For convenience we also record here the values of $\mathbf{a},\mathbf{b},\mathbf{c}$ and $\mathbf{d}$ of our map given in Eq.\;\eq{MSW2map}:
\begin{equation}
  \frac{v}{\kappa R}  ~=~   \hat \psi  \,, \quad~~  
	\psi = \frac{1}{2 }\,\hat \psi -   \frac{1}{\kappa R}\hat v \,,  \quad~~ (u,\phi) ~\, \mathrm{fixed}
	\quad~~ \Leftrightarrow \quad~~ \big(\mathbf{a},\mathbf{b},\mathbf{c},\mathbf{d}\big) ~=~ \left(0,\kappa, -\coeff{1}{\kappa }, \coeff{1}{2}\right) .  \label{MSW1map}
\end{equation}
While this is the particular map of interest to us, for much of what follows we shall derive expressions valid for general  $\mathbf{a},\mathbf{b},\mathbf{c}$ and $\mathbf{d}$.

\subsection{General spectral transformation of the metric functions} 
\label{sec:GenSTmetric}

The first, and simplest, step in computing the effect of spectral transformations is to start with the metric.  The structure of the argument closely follows that of \cite{Bena:2008wt}, however we use a more convenient formulation that may be found in  \cite{Niehoff:2013kia}.  

We first write the six-dimensional metric, (\ref{sixmet}), in terms of the double circle fibration and quantities that will remain invariant under spectral transformations:
\begin{equation}
ds_6^2 ~=~  - 2  H^{-1} \, (du +  \varpi )\, \alpha_1  ~+~  H^{-3} \,\big[ \cQ\, \alpha_1^2 ~+~ \alpha_2^2  \big] ~+~    H \, d \vec y \cdot d \vec y \,,  \label{newsixmet}
\end{equation}
where
\begin{equation}
\begin{aligned}
\alpha_1 &~\equiv~ V \, (dv + \beta) ~=~   V\, (dv +  \xi) ~+~ K_3 (d\psi + A)   \,, \\
\alpha_2  &~\equiv~  K_3^2 \,  \nu\,  (d\psi + A)  ~-~   V^2 \, \mu \, (dv +  \xi)  \,,
\end{aligned}
\label{alphasdefn}
\end{equation}
and where the functions $H$, $\cQ$ and $\nu$ are defined by:
\begin{equation}
 H   ~\equiv~  V \, \sqrt{\cP}   \,, \qquad \cQ ~\equiv~   -(\cP \cF \,V ~+~ \mu^2 V^2)\,, \qquad  \nu ~\equiv~ \frac{V}{K_3} \, \bigg( \frac{\cP}{K_3} - \mu\bigg)\,. \label{HQtmudefns}
\end{equation}

By definition, $u$ and the three-dimensional base parametrized by $\vec y$ are inert under the spectral transformation.  Since the overall form of the ansatz \eq{newsixmet} is required to remain invariant, it follows immediately   that $H$ and $\alpha_1$ are invariant since they multiply $d \vec y \cdot d \vec y$ and $du$.   Since $\alpha_1$ and $H$ are invariant, the invariance of the other terms that involve $\alpha_1$ and $H$ implies that $\varpi$, $\cQ$ and $\alpha_2^2$ must  be invariant.   {\it A priori} there could be a sign flip in the transformation of $\alpha_2$, but this is resolved by examining the transformations of the gauge fields, as we shall see shortly.  
 
We therefore find that, under the general spectral transformations \eq{modmap1}, we have:
\begin{equation}
\widehat{\alpha}_1~=~  \alpha_1 \,, \qquad  \widehat{\alpha}_2 ~=~  \alpha_2 \,, \qquad \widehat H~=~   H \,, \qquad   \widehat \cQ ~=~   \cQ \,, \qquad \widehat \varpi ~=~   \varpi \,.
\label{modinvs1}
\end{equation}
The invariance of $ \alpha_1$ implies:
\begin{equation}
V \;=\;  \mathbf{d} \, \widehat  V   -  \frac{\mathbf{c}}{R} \widehat K_3 \,, \qquad K_3 \;=\;  - \mathbf{b} R \,  \widehat  V +  \mathbf{a} \, \widehat  K_3\,, \qquad 
\xi  \;=\;  \mathbf{a} \, \widehat  \xi   +  \mathbf{b} R \, \widehat A \,, \qquad A \;=\;   \frac{\mathbf{c}}{R}  \,  \widehat  \xi + \mathbf{d} \, \widehat  A \,.\label{modtrf1}
\end{equation}
Similarly,  the invariance of $\alpha_2$ provides the transformations of $\mu$ and $\nu$:
\begin{equation}
 \nu\, K_3^2  ~=~  \mathbf{a} \,  (\hat {\nu}\, \widehat K_3^2)   ~+~  \mathbf{b} R \, (\hat \mu\, \widehat V^2)   \,, \qquad  \mu\, V^2  ~=~  \frac{\mathbf{c}}{R} \,  (\hat {\nu}\, \widehat K_3^2)   ~+~  \mathbf{d} \, (\hat \mu\, \widehat V^2) \,.\label{modtrf3}
\end{equation}
One can  use this and the expression for $\nu$ in (\ref{HQtmudefns})  to write 
\begin{equation}
\hat \mu  ~=~   \frac{V}{\widehat V}\, \mu  ~-~  \frac{\mathbf{c}}{R}  \,  \frac{ V}{\widehat V^2}\,\cP \,, \qquad  \mu  ~=~   \frac{ \widehat V}{ V}\,\hat \mu  ~+~   \frac{\mathbf{c}}{R} \,  \frac{ \widehat V}{ V^2}\,\widehat \cP  \,.\label{modtrf4}
\end{equation}
The transformation of $\cF$ then follows from the invariance of $\cQ$ and the other transformations:
\begin{equation}
\widehat \cF ~=~ \frac{\widehat V}{V}\,  \cF ~+~ 2\,\frac{\mathbf{c}}{R}\, \mu ~-~\frac{\mathbf{c}^2}{R^2} \, \cP \,  \widehat V^{-1} \,, \qquad
\cF ~=~ \frac{ V}{ \widehat V}\, \widehat \cF ~-~ 2\,\frac{\mathbf{c}}{R}\, \hat \mu ~-~ \frac{\mathbf{c}^2}{R^2}\, \widehat \cP \,   V^{-1}  \,.\label{modtrf5}
\end{equation}

Finally, we recall that $\cP$ is given by  (\ref{Pform0}) and that $H = V \sqrt{\cP}$ is invariant. Indeed, by examining ansatz for the tensor field strengths \eq{niceGform} in the next subsection, we shall see that each of the $V Z_a$ is separately invariant $(a=1,2,4)$.
An alternative way to see this is to note that if the $Z_a$ did transform into each other under spectral transformations then it would have been evident in the older, well-understood $v$-independent spectral transformations reviewed in Section \ref{sec:SFGT1}. However, in such intrinsically five-dimensional solutions, the individual $Z_a^{-1}$'s are electrostatic potentials of distinct fields and do not transform into one another.  Either way, we see that $V Z_a$ is invariant for each $a$ and so we have 
\begin{equation}
 V \;\! Z_a ~=~ \widehat  V \;\! \widehat Z_a \qquad \Rightarrow\qquad   Z_a
 ~=~  \frac{\widehat V}{ \mathbf{d} \, \widehat V  - \coeff{\mathbf{c}}{R}  \, \widehat K_3 } \, \widehat Z_a \,,
\qquad~~
a=1,2,4 
\,.
\label{modtrf2}
\end{equation}
%

\subsection{General spectral transformation of the gauge fields} 
\label{sec:GenSTgauge}

The last part of the spectral transformation that we will need is the transformations of the $\Theta^{(I)}$. 
Observe that the first ``electrostatic'' terms in the covariant version of the ansatz for the tensor gauge fields in  (\ref{G-ans-cov}) can be written as
\begin{equation}
- \frac{1}{2}\,\frac{\eta^{ab} Z_b}{\cal P}\,(du + \omega ) \wedge (dv + \beta)\,
~=~  - \frac{1}{2}\,\frac{(\eta^{ab} \;\! V Z_b) }{(V^2 \,\cal P)}\,(du + \omega ) \wedge \alpha_1 \,. \label{Pots1}
\end{equation}
From the invariance of $V \sqrt{\cP}$, $du$ and $\alpha_1$, it follows immediately that the $V Z_a$ are separately invariant under spectral transformations.  Going one step further, one can write the two-form in (\ref{Pots1}) in terms of $\alpha_1 \wedge \alpha_2$ and conclude that the $\alpha_2$ is invariant under spectral transformations and that there are, indeed, no sign changes.  This proves the invariance claims made in the previous subsection. 

In analyzing the transformations of the rest of these gauge fields it is useful to introduce the following operators:
\begin{equation}
\vec {\mathscr{D}} ~\equiv~ \vec \nabla ~-~   \vec A \,\partial_\psi ~-~   \vec \xi \,\partial_v \,,
\qquad\quad
\eth  ~\equiv~  V \, \partial_\psi ~-~   K_3 \,\partial_v 
  \,.\label{curlyD-fiberder}
\end{equation}
These operators are  invariant under spectral transformations.

Thus far, our discussion has been applicable to general BPS solutions.  In order to simplify the algebra in disentangling the gauge fields we will now make the further assumption, which underlies the broad class of solutions considered in this paper: namely that the four-dimensional base metric, $ds_4^2(\cB)$, and the vector field, $\beta$, are independent of $v$. 
The other ansatz quantities are allowed to depend on all the other coordinates (except $u$).

The last part of the spectral transformation can be extracted by noting that  the two-form
\begin{equation}
  d v \wedge d \psi ~=~   d \hat v \wedge d\hat\psi      \, \label{inv2form}
\end{equation}
is an invariant, and therefore the components of the tensor field strengths proportional to this two-form must also be invariant. 
We define one-forms, $\lambda^{(a)}$, on $\IR^3$ via:
\begin{eqnarray}
\Theta^{(a)} &\!\!\equiv\!\!& (1 + *_4)\;\! \big [(d\psi +A) \wedge \lambda^{(a)} \big] \,, 
\quad a~=~1,2 \; ;
\qquad
\thetafour ~\equiv~   (1 + *_4)\, \big [(d\psi +A) \wedge \lambdafour \big] \,. \qquad~~
\label{lamdefn1}  
\end{eqnarray}
Note that we have defined $\lambda^{(1)}$ and $\lambda^{(2)}$ with upstairs indices, and $\lambda_4$ with a downstairs index, corresponding to our conventions for the $\Theta$ quantities.
The component of $G^{(1)}$  proportional to $dv \wedge d\psi$ is 
\begin{equation}\label{dvdspipart}
 \frac{1}{2} \left[ \lambda^{(1)} \;+\; \mathscr{D} \left( \frac{Z_2}{\cal P} \, \mu \right) 
\; - \;\eth \left(  \frac{Z_2}{V\cal P} \;\! (du + \varpi )\right) 
\right] \wedge dv \wedge d\psi \,
\end{equation}
and a similar expression holds for $G^{(2)}$ and $G_4$ in terms of $\lambda^{(2)}$, $Z_1$ and $\lambdafour$, $Z_4$ respectively.
Since $du$, $\varpi$, $(Z_a V)$ and $\eth$ are all invariant under $SL(2,\mathds{Q})$ transformations, we see that 
\begin{equation}\label{oneforminv}
\lambda^{(1)} ~+~{\mathscr{D}} \left( \frac{Z_2 \;\! \mu}{V\cal P}  \right) ,  \qquad 
\lambda^{(2)} ~+~{\mathscr{D}} \left( \frac{Z_1 \;\! \mu}{V\cal P}  \right) ,  \qquad 
\lambdafour ~+~{\mathscr{D}} \left( \frac{Z_4 \;\! \mu}{V\cal P} \right)   \qquad 
\end{equation}
are invariant under $SL(2,\mathds{Q})$ transformations.  Since $\mathscr{D}$ is invariant, it follows that:
\begin{equation}\label{lamtrf1}
\hat \lambda^{(1)} \,=\, \lambda^{(1)} +\frac{\mathbf{c}}{R} {\mathscr{D}} (\medhat V^{-1} Z_2)   \,, \qquad 
\hat \lambda^{(2)} \,=\,  \lambda^{(2)} + \frac{\mathbf{c}}{R}  {\mathscr{D}} (\medhat V^{-1} Z_1)  \,, \qquad 
\lambdahatfour  \,=\,    \lambdafour + \frac{\mathbf{c}}{R}   {\mathscr{D}} (\medhat V^{-1} Z_4)\,,
\end{equation}
where we have used (\ref{modtrf4}).
From this, one obtains the transformed $\widehat \Theta^{(a)}$ using:
\begin{eqnarray}
\hspace{-7mm}\medhat\Theta^{(a)} &\!\equiv\!& (1 + \hat*_4)\, \big [(d\hat\psi +\medhat{A}) \wedge \hat\lambda^{(a)} \big] \,,  \quad a~=~1,2 \; ;
\qquad~~
\thetahatfour ~\equiv~  (1 + \hat*_4)\, \big [(d\hat\psi +\medhat{A}) \wedge \lambdahatfour \big] \,. \quad
\label{ThetaRes1}  
\end{eqnarray}
Note that the Hodge duality operations in Eqs.\;(\ref{lamdefn1}) and (\ref{ThetaRes1}) involve the respective (in general different) GH base metrics.

While the three-forms $G^{(a)}$ have several other components, their complete invariance under the spectral transformation follows from the invariance of (\ref{oneforminv}) and the transformation laws given in Section \ref{sec:GenSTmetric}.

\section{Constructing M-theory superstrata}
\label{sec:GeneralContracting}

We now review the construction of superstrata in the D1-D5 frame and then map these solutions across to the M-theory frame. 

\subsection{D1-D5-P Superstrata}
\label{ss:SSgeom}

The superstratum is obtained by adding momentum waves to the background of the circular supertube~\cite{Bena:2015bea,Bena:2016agb}.  
Currently the most general solutions to the first layer, (\ref{eqZTheta}), of the BPS system are known \cite{Shigemori:2013lta,Bena:2015bea,Bena:2016ypk}.  However not all the corresponding solutions to the second layer, (\ref{eqFomega}), are  known explicitly.  On the other hand, solutions based upon a single mode have been studied extensively \cite{Bena:2015bea,Bena:2016ypk} and complete solutions can be obtained through straightforward computations. In this section we will follow the same route and consider for concreteness the superstrata with a single excited mode constructed in~\cite{Bena:2016ypk}, generalized to $\kappa>1$ (see~\cite{Bena:2016agb} for a discussion).  It should be remembered that the BPS equations are linear and so extending to arbitrary superpositions of modes is more of a technical, rather than conceptual, issue\footnote{The difficulty lies in finding particular solutions of  (\ref{eqFomega}) in which the sources on the right-hand side come from products of two generic but different modes.}. 

In order to map to the M-theory frame, one must impose an isometry along the Hopf fiber of the $\IS^3$; our methods apply generally to any smooth solution with such an isometry. Since this isometry is only necessary for the final step of reducing to five dimensions, we will first work more generally, before eventually imposing the isometry in Section \ref{sec:Msuperstrata}.

The four-dimensional base metric remains the flat $\mR^4$ given in \eq{basemet}, which has the standard orthonormal frame 
\begin{equation}
e_1 ~=~ \frac{\Sigma^{1/2} }{(r^2 + a^2)^{1/2}} \, dr\,, \quad   e_2 ~=~\Sigma^{1/2}   \, d\theta\,, \quad   e_3 ~=~(r^2 + a^2)^{1/2}  \sin \theta  \, d\varphi_1\,, \quad   e_4 ~=~ r  \cos  \theta \, d\varphi_2 \,.
\label{frames1}
\end{equation} 
We define the self-dual two-forms $\Omega^{(1)}$, $\Omega^{(2)}$ and $\Omega^{(3)}$:
\begin{equation}\label{selfdualbasis}
\begin{aligned}
\Omega^{(1)} &\equiv \frac{dr\wedge d\theta}{(r^2+a^2)\cos\theta} + \frac{r\sin\theta}{\Sigma} d\varphi_1\wedge d\varphi_2 ~=~  \frac{1}{\Sigma \, (r^2+a^2)^\frac{1}{2}  \cos\theta} \,(e_1 \wedge e_2 +  e_3 \wedge e_4)\,,\\
\Omega^{(2)} &\equiv  \frac{r}{r^2+a^2} dr\wedge d\varphi_2 + \tan\theta\, d\theta\wedge d\varphi_1  ~=~  \frac{1}{\Sigma^\frac{1}{2}\, (r^2+a^2)^\frac{1}{2} \cos\theta} \,(e_1 \wedge e_4 +  e_2 \wedge e_3)   \,,\\
 \Omega^{(3)} &\equiv \frac{dr\wedge d\varphi_1}{r} - \cot\theta\, d\theta\wedge d\varphi_2~=~  \frac{1}{\Sigma^\frac{1}{2}\, r \sin\theta} \,(e_1 \wedge e_3 -  e_2 \wedge e_4)  \,.
\end{aligned}
\end{equation}
We take the one-form $\beta$ to be that of a supertube, given in \eq{eq:betaomegaST}; the two-form $\Theta^{(3)}=d\beta$ is then given by
\begin{equation} 
\Theta^{(3)}
~=~\frac{2 \kappa R_y a^2 }{\Sigma^2} \, ( (r^2 + a^2)\cos^2 \theta \,  \Omega^{(2)}   - r^2 \sin^2 \theta \,  \Omega^{(3)} ) \,.
\label{Theta3form1}
\end{equation} 
The solutions have a non-trivial phase dependence that is parameterized by a positive integer $k$ and non-negative integers $m$, $n$ subject to $m\le k$, and that takes the form of the combination
\begin{equation}
\label{SSmodes1}
\chi_{k,m,n} ~\,\equiv\,~ (m+n) \coeff{ v}{\kappa R_y}  +  (k-m) \;\! \varphi_1 -  m \;\! \varphi_2   
~\,=~\, (m+n) \,\coeff{ v}{\kappa R_y} + \coeff{1}{2}(k-2m) \, \psi - \coeff{k}{2}\;\! \phi   \,.
\end{equation}
In order to have a single-valued supergravity solution in this frame, one must take $(m+n)$ to be a multiple of $\kappa$. We will assume this for all solutions in the D1-D5-P frame. A more detailed discussion of this point can be found in~\cite{Bena:2016agb}.

We also define:
\begin{equation}\label{Deltadefn}
\Delta_{k,m,n}~\equiv~ \frac{a^k \, r^n }{(r^2+a^2)^{\frac{k+n}{2} }}\,\sin^{k-m}\theta\,\cos^m\theta\,.
\end{equation}
As discussed earlier,   we work in the parameterization in which $u$ is the natural time coordinate in five dimensions, such that $\beta$ and the $\Theta^{(I)}$ are rescaled with respect to the conventions of \cite{Bena:2015bea,Bena:2016ypk}, as discussed in Appendix \ref{app:supertube}. With this in mind, we take $\beta$ and $ds_4^2$ to be as given in Eqs.\;\eq{eq:betaomegaST} and \eq{GHmet2}, and consider the following solution \cite{Bena:2016ypk} to the first layer of BPS equations \eq{eqZTheta}:
\begin{equation}
\begin{aligned}
Z_1  ~=~ & \frac{Q_1}{\Sigma} \left( 1 + \frac{b_4^2}{2 a^2+b^2} \;\! \Delta_{2k,2m,2n}  \;\! \cos \chi_{2k,2m,2n} \right) \,,
\qquad Z_2  ~=~ \frac{Q_5}{\Sigma}  \,, \qquad\Theta^{(1)}  ~=~ 0 \,,  \cr
Z_4  ~=~ &   b_4 \frac{\kappa R_y}{\Sigma} \;\! \Delta_{k,m,n}   \;\! \cos \chi_{k,m,n} \,, \\
\Theta^{(2)}  ~=~ & {}-b_4^2 \;\!\frac{\kappa R_y}{2Q_5}  \;\!  \Delta_{2k,2m,2n} \, \Big[\Big(2(m+n) \, r\, \sin \theta + 2n\, \Big(\frac{m}{k} -1\Big)\,  \frac{\Sigma}{r \, \sin \theta} \Big)\, \sin \chi_{2k,2m,2n} \,\Omega^{(1)} \\
&\qquad \qquad \qquad\qquad ~+~ \cos  \chi_{2k,2m,2n} \, \Big(2m\, \Big(\frac{n}{k} + 1\Big)\, \Omega^{(2)} ~+~  2n\,\Big(\frac{m}{k} - 1\Big)\, \Omega^{(3)}\Big)\Big]  \,, \cr
\thetafour  ~=~ & {} b_4 \;\!  \Delta_{k,m,n} \, \Big[\Big((m+n) \, r\, \sin \theta + n\, \Big(\frac{m}{k} -1\Big)\,  \frac{\Sigma}{r \, \sin \theta} \Big)\, \sin \chi_{k,m,n} \,\Omega^{(1)} \\
&\qquad \qquad \qquad\qquad ~+~ \cos  \chi_{k,m,n} \, \Big(m\, \Big(\frac{n}{k} + 1\Big)\, \Omega^{(2)} ~+~  n\,\Big(\frac{m}{k} - 1\Big)\, \Omega^{(3)}\Big)\Big] \,.
\end{aligned}
\label{ZThetaform1}
\end{equation}
Note that there is no summation over $(k,m,n)$ and that we have chosen the Fourier modes in $(Z_1,\Theta^{(2)})$ to be related to those of $(Z_4,\thetafour)$.  This is the now-standard ``coiffuring procedure'' often used to obtain regular solutions \cite{Mathur:2013nja,Bena:2013ora,Bena:2015bea,Bena:2016ypk}.  
In more detail, the $(Z_4,\thetafour)$ system contains mode-numbers $(k,m,n)$ with Fourier coefficient $b_4$, and the $(Z_1,\Theta^{(2)})$ system contains mode-numbers $(2k,2m,2n)$ with Fourier coefficient $b_4^2$.
For the simple solutions considered here, this means that the metric in fact has isometries along $v$, $\psi$, $\phi$, even though the isometries are broken by the tensor fields. (In more general superstratum solutions, only the null isometry along $u$ will be present in the metric). In particular, the oscillatory modes cancel in the metric function:
\begin{equation}
\cP ~=~   Z_1\;\! Z_2 \,-\, Z_4^2 \,. \label{Pform2}
\end{equation}
There are, however, `RMS terms', proportional to $b_4^2$, that survive in $\cP$ and in other parts of the metric.  

The charges and amplitudes of the oscillations are related through the following regularity constraint:
\begin{equation}
\frac{Q_1Q_5}{\kappa^2 R_y^2}  ~=~   a^2 +  \frac{b^2}{2}  \,,  \qquad\qquad 
b^2 ~\equiv~ \left[ {k \choose m}{k+n-1 \choose n} \right]^{-1}b_4^2 \,.
  \label{eq:regularity-b-b4}
\end{equation}
For $b=b_4=0$, this gives the radius relation \eq{streg} that emerges from the requirement  that the metric for the unexcited supertube be non-singular, up to the same $\mZ_{\kappa}$ orbifold singularity at ($r=0,\theta=\pi/2$) discussed around Eq.\;\eq{eq:lattice-D1D5}.

One can also introduce frames based upon the GH form of the metric:  
\begin{align}
\tilde e_1 ~=~& \frac{(\Lambda \, \Sigma)^{1/2}}{2\, (\Sigma - \Lambda)^{1/2}} \,(d  \psi +  A) \,, \qquad   \tilde e_2 ~=~\frac{(\Sigma - \Lambda)^{1/2}}{(r^2+a^2)^{1/2} }   \, dr\,, \\
\tilde e_3 ~=~ &(\Sigma - \Lambda)^{1/2}  \, d \theta \,, \quad   \tilde e_4 ~=~\frac{(\Sigma - \Lambda)^{1/2}}{(\Lambda \, \Sigma)^{1/2}}  \, r \, (r^2+a^2)^{1/2} \,\sin \theta \,   \cos \theta \,  d\phi \,,
\label{frames2}
\end{align} 
and the standard self-dual two-forms, $\widetilde \Omega^{(I)}$:
\begin{equation}\label{tildeselfdualbasis}
\widetilde \Omega^{(1)} ~\equiv~  \tilde e_1 \wedge \tilde e_2 +  \tilde e_3 \wedge \tilde e_4\,,\quad 
\widetilde\Omega^{(2)} ~\equiv~  \tilde e_1 \wedge \tilde e_3 -   \tilde e_2 \wedge \tilde e_4  \,,\quad
\widetilde \Omega^{(3)} ~\equiv~   \tilde e_1 \wedge \tilde e_4  +  \tilde e_2 \wedge \tilde e_3  \,. 
\end{equation}
Note that with the transformation of coordinates (\ref{coords1}), the orientation and dualities of the $\tilde e_a$  match those of the $e_a$:  $e_1 \wedge \dots\wedge e_4 =  \tilde e_1 \wedge \dots\wedge \tilde e_4$.

\subsection{Transforming the superstrata}
\label{ss:Dgeom}

In Section \ref{sec:supertube} we showed how to get to the M-theory frame by making the following coordinate transformation on the class of solutions described in Section \ref{ss:SSgeom}: 
\begin{equation}
  \frac{v}{\kappa R}  ~=~   \hat \psi  \,, \quad~~  
	\psi = \frac{1}{2 }\,\hat \psi -   \frac{1}{\kappa R}\hat v \,,  \quad~~ (u,\phi) ~\, \mathrm{fixed}
	\quad~~ \Leftrightarrow \quad~~ \big(\mathbf{a},\mathbf{b},\mathbf{c},\mathbf{d}\big) ~=~ \left(0,\kappa, -\coeff{1}{\kappa }, \coeff{1}{2}\right) .  \label{MSW1map-2}
\end{equation}
Under this mapping, the phase dependence, given in (\ref{SSmodes1}), becomes 
\begin{equation}
\label{SSmodestrf}
\hat\chi_{k,m,n} ~=~ - (k-2m) \coeff{\hat v}{\kappa R_y} + \coeff{1}{2}\left( n+ \coeff{k}{2} \right) \hat\psi - \coeff{k}{2} \phi \,.
\end{equation}
The lattice of identifications is re-declared to be \eq{vhat-psihat-periods-2}, so single-valuedness requires that $(k-2m)$ is a multiple of $\kappa$, and that $k$ is even (recall that $k$ is by definition a positive integer). We will eventually impose an isometry along $\hat{v}$, by setting $k=2m$, however for most of the following we shall keep both $k$ and $m$ in the analysis.

\subsubsectionmod{Transforming the metric quantities}
\label{sec:metrictransf}

The six-dimensional metric is now expressed in the parameterization:
\begin{equation}
ds_6^2 ~=~    -\frac{2}{\sqrt{\widehat \cP}} \, (d\hat v+\widehat \beta) \big(du + \widehat \omega + \tfrac{1}{2}\, \widehat  {\mathcal{F}} \, (d\hat v+\widehat \beta)\big) 
~+~  \sqrt{\widehat \cP} \, \widehat{ds}_4^2(\cB)\,.  \label{sixmethat}
\end{equation}
Using the transformation rules \eq{modtrf1}, the transformed functions and vector fields in the fibrations are: 
\begin{eqnarray}
  \widehat V ~=~ \frac{1}{r_+} -  \frac{1}{r_-}  &=& \frac{4\;\!(\Sigma-\Lambda)}{\Lambda \, \Sigma}  \,,  
	\qquad\quad
 \widehat A   ~=~  - \frac{2 \;\! a^2 \;\! (\Sigma+ \Lambda ) }{\Lambda \, \Sigma} \, \sin^2 \theta \,   \cos^2 \theta\,  d \phi \,,   \qquad \label{MSWmap1} \\
\widehat{K}_3  ~=~   \frac{\kappa R}{2} \left(\frac{1}{r_+} +   \frac{1}{r_-} \right)  
&=& \kappa R_y \frac{(\Sigma+\Lambda)}{\Lambda \, \Sigma}
\,, 
  \qquad ~~
	\widehat \xi  ~=~ - \kappa R_y \;\! \frac{ (\Sigma-  \Lambda ) }{2\;\! a^2 \;\! \Lambda \;\! \Sigma} \, r^2 \,  (r^2+a^2) \, d \phi \,.  \label{MSWmap2}  
\end{eqnarray}
The vector field, $\widehat \beta$ is then given by:
\begin{equation}
\medhat \beta ~\,=\,~    \frac{\widehat K_3}{\widehat V}\, (d \hat\psi +\widehat A)~+~ \widehat \xi  
~\,=\,~  \frac{\kappa R_y}{2\cos 2\theta }\,\Big(\frac{ (2 \,r^2+a^2)}{2\, a^2}\, d \hat\psi - d \phi  \Big) \,.  \label{betaform3}
\end{equation}
One can also check that the two-form $\Theta^{(3)}$ is given by:
\begin{equation} 
\widehat \Theta^{(3)} ~=~ d \medhat \beta ~=~
 - (1+ \hat *_4) \, \Big( \frac{\kappa R_y}{4}\, (d \hat \psi +\medhat A)\wedge  \Big[ d\,\Big(\frac{\Sigma + \Lambda}{\Sigma - \Lambda} \Big)\Big] \Big) \,,
\label{trfTheta3}
\end{equation} 
which matches the classic, five-dimensional form  \cite{Bena:2005va,Berglund:2005vb,Bena:2007kg}.

The four-dimensional metric becomes:
\begin{equation}
\widehat{ds}_4^2 ~=~ \frac{\Lambda \, \Sigma}{4\, (\Sigma - \Lambda)} \,(d \hat \psi + \medhat A)^2 ~+~ (\Sigma -\Lambda)\, \bigg( \frac{dr^2}{(r^2 + a^2)} ~+~ d \theta^2 ~+~ \frac{ r^2 \,  (r^2+a^2) }{\Lambda \, \Sigma} \,   \sin^2 \theta \,   \cos^2 \theta \, d \phi^2 \bigg)   \,.
\label{bipolmet}
\end{equation} 
We also introduce frames based upon the GH form of the metric:  
\begin{align}
\hat e_1 ~=~& \frac{(\Lambda \, \Sigma)^{1/2}}{2\, (\Sigma - \Lambda)^{1/2}} \,(d \hat \psi +\medhat A) \,, \qquad   \hat e_2 ~=~\frac{(\Sigma - \Lambda)^{1/2}}{(r^2+a^2)^{1/2} }   \, dr\,, \\
\hat e_3 ~=~ &(\Sigma - \Lambda)^{1/2}  \, d \theta \,, \quad   \hat e_4 ~=~\frac{(\Sigma - \Lambda)^{1/2}}{(\Lambda \, \Sigma)^{1/2}}  \, r \, (r^2+a^2)^{1/2} \,\sin \theta \,   \cos \theta \,  d\phi \,,
\label{frames3}
\end{align} 
and the standard self-dual two-forms, $\widehat \Omega^{(I)}$:
\begin{equation}\label{selfdualbasishat}
\widehat \Omega^{(1)} ~\equiv~  \hat e_1 \wedge \hat e_2 +  \hat e_3 \wedge \hat e_4\,,\quad 
\widehat\Omega^{(2)} ~\equiv~  \hat e_1 \wedge \hat e_3 -   \hat e_2 \wedge \hat e_4  \,,\quad
\widehat \Omega^{(3)} ~\equiv~   \hat e_1 \wedge \hat e_4  +  \hat e_2 \wedge \hat e_3  \,.
\end{equation}

Note that $\Sigma - \Lambda = a^2 \cos 2 \theta$ vanishes at $\theta = \pi/4$.  The base metric and the forms are thus singular on this locus.  This is a standard feature of using an ambi-polar base on which $\widehat V$ vanishes.  The complete physical fields are, of course, completely smooth because locally we have simply made a coordinate transformation of a smooth solution.

One can now perform the spectral transformation on the metric functions \eq{modtrf2} to obtain:
\begin{equation}
\widehat  Z_1 ~=~   \frac{Q_1}{\Sigma -   \Lambda}~+~ \widehat  Z_{1}^{\mathrm{(osc)}} \,, \qquad 
\widehat  Z_{1}^{\mathrm{(osc)}} ~\equiv~
 \frac{Q_1}{\Sigma -   \Lambda} \;\! \frac{b_4^2}{2a^2+b^2}\;\! \Delta_{2k,2m,2n} \;\! \cos \hat \chi_{2k,2m,2n} \,, \qquad Z_2   ~=~ \frac{Q_5}{\Sigma - \Lambda} \label{spectransf1}\, .
\end{equation}
Observe that the phase dependence has been converted to $\hat \chi_{2k,2m,2n}$.

\subsubsectionmod{Transforming the fluxes}
\label{sec:fluxtransf}

To transform the magnetic fluxes we now extract the one-form, $\lambda^{(2)}$, defined in (\ref{lamdefn1}) by taking the coefficient of $d\psi = (d \varphi_1 + d \varphi_2)$ in $\Theta^{(2)}$.  We find:
\begin{equation}\label{lam2form1}
\begin{aligned}
\lambda^{(2)} ~=~ 
 b_4^2 \;\!\frac{\kappa R_y}{2Q_5} \frac{1}{k}  \, \Delta_{2k,2m,2n} \,  \Big[ \,  & r\, \Big( \frac{m(k+n)}{r^2+a^2} - \frac{n(k-m)}{r^2} \Big) \,\cos \chi_{2k,2m,2n}   \, dr    \\
  &+  \big( n(k-m) \;\! \cot \theta  + m (k+n)  \;\! \tan \theta  \big) \,\cos \chi_{2k,2m,2n}   \, d\theta  \\
  &+  \Big( n(k-m)  - k(m+n) \frac{r^2 \sin^2 \theta }{\Sigma} \Big)  \,\sin \chi_{2k,2m,2n}  \, d\phi  \, \Big] \,.
\end{aligned}
\end{equation}
Using (\ref{lamtrf1}) with $\mathbf{c}=-\coeff{1}{\kappa}$, we have
\begin{equation}\label{lamtrf2}
\hat \lambda^{(2)} \,=\,  \lambda^{(2)} - \frac{1}{\kappa R}  {\mathscr{D}} (\medhat V^{-1} Z_1)   \, ,
\end{equation}
which leads to: 
\begin{eqnarray}
\hat \lambda^{(2)} &=& -\frac{ 1}{4\kappa R_y} \,d_{(3)} \big[ (\Sigma + \Lambda) \, \widehat Z_1\big] 
~-~ \frac{a^2 \,\widehat Z_{1}^{\mathrm{(osc)}}}{2 \kappa  R_y} \frac{1}{r \;\! (r^2 +a^2)} \frac{1}{k}\, (k-2m)\big(a^2 n + (k+2n) r^2 \big) \cos 2 \theta \, dr    \cr
  &&\qquad\qquad\quad ~+~ 
	\frac{a^2 \, \widehat Z_{1}^{\mathrm{(osc)}}}{4\kappa  R_y}\frac{1}{\sin  \theta\, \cos  \theta}\frac{1}{k}(k+2n)\big(k + (k-2m) \cos 2 \theta \big)\, \cos 2 \theta \, d\theta  \cr
  &&\qquad\qquad\quad ~+~    \frac{ \partial_\phi  \widehat Z_{1}^{\mathrm{(osc)}}}{4 \kappa R_y} \,\frac{1}{k^2}
\big(a^2 k(k+2n)  \;\!+\;\!  2(k-2m)(a^2 n - k r^2 ) \cos 2 \theta  \big) \, d\phi 
	\,,
\label{lam2form0}
\end{eqnarray}
where $d_{(3)}$ is the exterior derivative on the $\IR^3$ of the GH space.

One can then find $\widehat \Theta^{(2)}$ using \eq{ThetaRes1}, whereupon one can explicitly verify that $\widehat Z_1$ and $\widehat\Theta^{(2)}$ satisfy the first layer of the equations: 
 \begin{equation}\label{eqZ1Theta2a}
 \hat*_4  \medhat D (\partial_{\hat v} \widehat Z_1)~-~   \medhat D\widehat \Theta^{(2)} ~=~ 0 \,,\quad  \medhat D \hat*_4  \medhat D \widehat Z_1~=~ -\widehat \Theta^{(2)}\wedge d\medhat\beta\,,\quad \widehat\Theta^{(2)}=  \hat*_4 \widehat\Theta^{(2)}\,,
 \end{equation}
 where
\begin{equation}
\medhat D~\equiv~ \hat d_{(4)} - \medhat \beta\wedge \frac{\partial}{\partial \hat v}\,, 
\end{equation}
where in turn $\hat d_{(4)}$ is the exterior derivative on the transformed four-dimensional base.
The remaining fluxes $\widehat \Theta^{(1)}$ and $\thetahatfour$ may be obtained similarly.

\subsubsectionmod{M-theory superstrata}
\label{sec:Msuperstrata}

To reduce to five dimensions we require solutions that are independent of the $\hat v$-fiber.   This means restricting the modes to those with $k=2m$, and hence with a phase, $\hat \chi_{k,m,n}$, in (\ref{SSmodestrf}) given by:
\begin{equation}
\hat \chi_{2m,m,n}  ~=~ \coeff{1}{2}  \, (n + m) \, \hat \psi ~-~m \,\phi  \,. \label{modered1}
\end{equation}
The expression for $\hat \lambda^{(2)}$ then simplifies significantly:
\begin{equation}\label{lam2form2}
\hat \lambda^{(2)} ~=~ 
 -\frac{ 1}{4\kappa R_y} \,d_{(3)} \big[ (\Sigma + \Lambda) \,\widehat Z_1\big]  
 ~+~ \frac{ a^2  (m+n)}{4\kappa R_y} \,\Big[  4\, \widehat Z_{1}^{\mathrm{(osc)}}  \, \cot 2 \theta \, d\theta  
~+~   \coeff{1}{m} \, \partial_\phi \widehat Z_{1}^{\mathrm{(osc)}} \, d\phi  \, \Big]  \,.
\end{equation}
The BPS equations also reduce to their five-dimensional form and, in particular, (\ref{eqZ1Theta2a}) implies that 
 $\widehat\Theta^{(2)}$, is self-dual and closed: 
 \begin{equation}\label{Theta2harm}
  \hat d \;\! \widehat \Theta^{(2)} ~=~ 0 \,,\qquad   \widehat\Theta^{(2)}=  \hat*_4 \widehat\Theta^{(2)}\,,
 \end{equation}
 and is thus ``harmonic'' on the GH base.     One can explicitly verify this using (\ref{lam2form2}) and (\ref{ThetaRes1}).
 
The harmonic forms on a standard Riemannian GH base are well-known (see, for example,  \cite{Bena:2007kg}).  For $\nc$ GH centers there are $(\nc-1)$ independent, smooth harmonic forms given by  the expressions in (\ref{harmfns}).  In particular, these harmonic forms are independent of the angles $(\psi, \phi)$.  It may therefore seem surprising that there is, in fact, a doubly infinite family of ``harmonic forms'' emerging from our solutions.  However, this is because the base is ambi-polar and hence singular on the locus $\widehat V = 0$.  This singular locus enables the ``harmonic forms'' to have (singular) sources on this locus and thus the system admits large families of solutions with oscillating magnetic fluxes.  Again, as with everything else in the ambi-polar formulation, the physical field strengths must be smooth.  In this paper smoothness is guaranteed because we derived the solution by a coordinate change of a smooth six-dimensional solution.  

Henceforth we will use the term {\it pseudo-harmonic forms} to refer to the generalized  ``harmonic forms'' that are singular on the degeneration locus ($\widehat V = 0$) of an ambi-polar geometry, and yet give rise to smooth physical fields in the complete solution.

The first analyses of five-dimensional BPS solutions were done over a decade ago  \cite{Bena:2005va,Berglund:2005vb,Bena:2007kg} and the pseudo-harmonic forms were missed in that analysis.  Given the ambi-polar structure of the base, many people were aware that the singular locus could allow the presence of new sources  that could generalize the usual known solutions.  The problem was that there was a vast range of singular sources available and no obvious systematic way to find precisely those sources that would lead to smooth physical field strengths.   That is, the possibility, let alone the classification, of non-trivial pseudo-harmonic forms remained unclear.  It is interesting to note that the possibility of ambi-polar metrics was first found by Giusto and Mathur \cite{Giusto:2004kj} by studying spectral flows of smooth supertube geometries.  In this paper we have used more general spectral transformations to discover precisely how to go beyond the standard analogs of Riemannian harmonic forms in five dimensions to obtain (hopefully complete\footnote{Complete here means the complete family of pseudo-harmonic forms within the terms of our definition.  Specifically, while singular on the degeneration locus of ambi-polar geometries,  pseudo-harmonic forms are required to lead to smooth BPS solutions in five dimensions.}) families of  pseudo-harmonic forms on our specific ambi-polar geometry.  It would be very interesting to see how  pseudo-harmonic forms might be  characterized, in terms of the differential geometry, and then computed for generic ambi-polar hyper-K\"ahler metrics.

The bottom line is that we have obtained a huge class of pseudo-harmonic forms and these lead to new families of smooth five-dimensional solutions with fluctuating fluxes.  As we have argued above, these solutions must be dual to microstates of the MSW string.

\section{An explicit example}
\label{sec:example}

We now give a complete explicit example.  It is one of the family of solutions discussed in the previous section and has parameters $(k,m,n)=(2,1,n)$. Since $k=2m$, this can be dualized to a smooth five-dimensional solution.

\subsection{The D1-D5-P superstrata}
\label{ss:superstratum}

 The quantities $\beta$ and $ds_4^2$ are again as given in Eqs.\;\eq{eq:betaomegaST} and \eq{GHmet2}. The quantities of the first layer of the BPS equations are as given in \eq{ZThetaform1} with $(k,m,n)=(2,1,n)$, for  a non-negative integer, $n$, where $(n+1)$ is a multiple of $\kappa$. The phase dependence of this solution is:
\begin{equation}
\chi_{2,1,n} ~=~ (n+1) \,\coeff{ v}{\kappa R_y} -  \phi   \qquad \Rightarrow \qquad
\hat\chi_{2,1,n} ~=~ \coeff{1}{2}\left( n+1 \right) \hat\psi - \phi \,.
\end{equation}
The relation between $b$ and $b_4$ required for regularity is 
\be
b^2 ~=~ \frac{b_4^2}{2(n+1)} \,.
  \label{eq:regularity-b-b4-21n}
\ee
Again generalizing the solution of~\cite{Bena:2016ypk} to $\kappa>1$, the solution to the second layer of BPS equations is:
\begin{align}
\cF ~=~ & -1 - \frac{b^2 }{2 \;\! a^2} +\frac{b_4^2 }{2\;\!a^2} \, 
 \frac{\Delta_{4,2,2n}}{\sin^2 \theta\cos^2 \theta}\,    \left( \frac{\Sigma}{4 \;\! a^2} 
+ \frac{1}{2(n+1)}\frac{r^2 (r^2 +a^2)}{a^4}  \right) \,, \cr 
\omega ~=~ & \omega_1 \;\! d\varphi_1 + \omega_2 \;\! d \varphi_2 \,, \label{eq:SS-layer2} \\ 
\omega_1 ~=~ & \frac{\kappa R_y}{\Sigma} \left(a^2 + \coeff{b^2 }{2 } \right) \sin^2 \theta
-\frac{b_4^2\;\! \kappa  R_y}{4 \;\! \Sigma}  \;\! \Delta_{4,2,2n}  \;\! \frac{r^2+a^2}{a^2}
\left( 1+ \frac{1}{2\;\!(n+1)} \frac{r^2}{a^2 \cos^2 \theta}\right) \,, \cr 
\omega_2 ~=~ &\frac{b_4^2\;\! \kappa  R_y}{4\;\!\Sigma}  \;\!  \Delta_{4,2,2n} \;\!
\frac{r^2}{a^2}\left( 1 + \frac{1}{2\;\!(n+1)} \frac{r^2 +a^2}{a^2 \,\sin^2 \theta}  
 \right) .
\nonumber
\end{align}
We record here the values of the other ansatz quantities that will be used when mapping to the M-theory frame:
\begin{align}
\cP  ~=~ & \frac{Q_1 Q_5}{\Sigma^2} \left( 1 \, - \,
\frac{b_4^2}{2a^2+b^2}\;\! \Delta_{4,2,2n} \right) , \cr
\mu ~=~ & \frac{\kappa R_y}{2\;\! \Sigma} \left(a^2 + \tfrac{b^2 }{2 } \right) \sin^2 \theta
- \frac{b_4^2}{2}\frac{\kappa R_y}{\Sigma} \;\! \Delta_{4,2,2n} 
\left( \frac{1}{4} - \frac{1}{2(n+1)} \frac{r^2(r^2 + a^2)}{a^4} \frac{\cot 2\theta}{\sin 2\theta} \right) ,
\label{eq:SS-ansatz-fns} \\
\varpi~=~ & \left[-  \! \left(a^2 + \frac{b^2 }{2 } \right) \! \frac{\kappa R_y}{\Sigma} \frac{r^2}{\Lambda} \sin^2\theta \cos^2 \theta \, 
+ \frac{b_4^2}{4}  \frac{\kappa R_y}{\Sigma} \frac{r^2}{\Lambda} \;\! \Delta_{4,2,2n} 
\frac{r^2+a^2}{a^2} \left( 1+ \frac{1}{2(n+1)} \Big( \;\!\! 1+ \frac{2 \;\! r^2}{a^2} \Big) \! \right) \right] d\phi \,.
\nonumber
\end{align}

These quantities lead to a family of smooth, CTC-free solutions, due to the coiffuring ansatz and appropriate choices of homogeneous solutions to the BPS equations~\cite{Bena:2016ypk}.

\subsection{The M-theory superstrata}
\label{ss:decon}
 
To transform to the M-theory frame we convert the base metric to GH form and transform the ansatz quantities recorded above. Using  (\ref{modtrf2}) and \eq{MSW1map},  the metric functions in the M-theory frame are
\begin{align}
\widehat Z_{1} ~\equiv~ & \, \frac{Q_1}{\Sigma - \Lambda} +  Z_{1}^{\mathrm{(osc)}} 
~=~    \frac{Q_1}{\Sigma - \Lambda} \left( 1 + \frac{b_4^2}{2 \;\! a^2+b^2} \,
\Delta_{4,2,2n} \;\! \cos \hat \chi_{4,2,2n} \right) \,, \cr
\widehat Z_{2}~=~ & \,  \frac{Q_5}{\Sigma - \Lambda}  \,, \qquad\qquad
\widehat Z_{4} ~=~  \frac{b_4  \kappa  R_y}{\Sigma - \Lambda}\,\Delta_{2,1,n} \;\! \cos \hat \chi_{2,1,n}   \,, \label{hatZI-P} \\
\widehat \cP  ~=~ & \, \frac{Q_1 Q_5}{(\Sigma - \Lambda)^2} \left( 1 \,-\, \frac{b_4^2}{2 \;\! a^2+b^2} \Delta_{4,2,2n} \right) \,.
\nonumber
\end{align}
The one-forms, $\hat \lambda^{(I)}$, are obtained from (\ref{lamtrf1})  with $\mathbf{c}=-\tfrac{1}{\kappa}$,
\begin{equation}\label{hatlamdas}
\hat \lambda^{(1)} \,=\,  -\frac{1}{\kappa R} {\mathscr{D}} (\medhat V^{-1} Z_2)   \,, \qquad 
\hat \lambda^{(2)} \,=\,  \lambda^{(2)} -\frac{1}{\kappa R}  {\mathscr{D}} (\medhat V^{-1} Z_1)  \,, \qquad 
\lambdahatfour  \,=\,   \lambdafour - \frac{1}{\kappa R}   {\mathscr{D}} (\medhat V^{-1} Z_4)
\end{equation}
where we have used that $\Theta^{(1)}=0$ from \eq{ZThetaform1}. We find
\begin{align}
\hat \lambda^{(1)}   ~=~ & -   d_{(3)} \bigg[\frac{(\Sigma + \Lambda) \, \widehat Z_2}{4 \kappa R_y}   \bigg]\,, \cr
\hat \lambda^{(2)} ~=~ &  -d_{(3)} \bigg[\frac{(\Sigma + \Lambda) \, \widehat Z_1}{4 \kappa R_y}   \bigg] ~+~ \frac{(n+1) \, a^2 }{4 \kappa R_y} \,\Big[4 \cot 2\theta \, \widehat Z_{1}^{\mathrm{(osc)}} \, d\theta ~+~ \big(\partial_\phi \widehat Z_{1}^{\mathrm{(osc)}} \big) \, d\phi \,  \Big]  \,, \label{lamform3} \\
\lambdahatfour   ~=~ &   - d_{(3)} \bigg[\frac{(\Sigma + \Lambda) \, \widehat Z_4}{4 \kappa R_y}   \bigg] ~+~ \frac{(n+1) \, a^2 }{4 \kappa R_y} \,\Big[2 \cot 2\theta \, \widehat Z_{4}\, d\theta ~+~ \big(\partial_\phi \widehat Z_{4}\big) \, d\phi \, \Big]  \,,
\nonumber
\end{align}
where we recall our notation that $d_{(3)}$ is the exterior derivative on the $\IR^3$ base of the GH space.

From the  above expressions one obtains the $\widehat \Theta^{(a)}$ using \eq{ThetaRes1}.
These are manifestly self-dual, and it is straightforward to verify that they are indeed closed.

The transformations that yield the last layer of the BPS system are (\ref{modinvs1}),  (\ref{modtrf4}) and (\ref{modtrf5}).  Using these  with $\mathbf{c}=-\tfrac{1}{\kappa}$, we obtain:
\begin{align}
\widehat \cF  ~=~ &
- \frac{1}{\Sigma-\Lambda} \, \left(  a^2 +\frac{b^2}{2}  - \frac{b_4^2}{2} \frac{\Delta_{4,2,2n}}{\sin^2 2 \theta}  \right) \,, \cr
\hat \mu  ~=~ &  \frac{\kappa R_y}{2(\Sigma-\Lambda)} 
\left[ \left( a^2 + \tfrac{b^2}{2} \right)\left(\frac{\Sigma}{\Sigma-\Lambda} - \cos^2\theta \right) \right.
\label{eq:hat-layer-2} \\
& \qquad\qquad\qquad + \left.
\frac{b_4^2}{a^2} \frac{\Delta_{4,2,2n}}{\sin 2\theta} 
\left(  \frac{2r^2(r^2+a^2)}{1+n}\cot 2\theta - a^2 (2r^2+a^2) \tan 2\theta \right)
\right]
\,, \cr
\medhat \varpi ~=~ & \varpi   \,.
\nonumber
\end{align}
One can then verify that these quantities, together with the hatted quantities and ansatz given in (\ref{sixmethat})--\eq{bipolmet}, do indeed satisfy the last layer of BPS equations for $k=2$ and $m=1$.  

Smoothness in five dimensions requires that $\hat \mu$ and the $\widehat Z_I$ are finite at the GH points while the absence of CTC's requires that $\hat \mu$ vanishes at the GH points.   The  GH points lie at $(r,\theta)$ = $(0,0)$ and $(r,\theta)$ = $(0,\pi/2)$ and if one sets $r=0$ in  (\ref{hatZI-P}) and  (\ref{eq:hat-layer-2}), one has: 
\begin{equation}
\label{Zmudisk}
\begin{aligned}
\widehat Z_1 & ~=~  \frac{Q_1}{a^2 \, \cos 2 \theta} \,, \qquad \widehat Z_2  ~=~  \frac{Q_5}{a^2 \, \cos 2 \theta} \,, \qquad \widehat Z_4  ~=~ 0  \,,  \\
\widehat Z_3 & ~=~ -   \widehat \cF   ~=~  \frac{1}{a^2 \, \cos 2 \theta} \, \left(  a^2 +\tfrac{b^2}{2}\right)  \,, \qquad \hat \mu  ~=~  \frac{\kappa R_y}{4\,a^2}\, \left(  a^2 +\tfrac{b^2}{2}\right)  \, \tan^2 2 \theta \,.
\end{aligned}
\end{equation}
A complete analysis of the global absence of CTCs is in general a difficult problem, often relying on numerical tests, and is beyond the scope of this paper. Here we content ourselves with observing that the five-dimensional solution satisfies the requisite local conditions, providing evidence that the spectral transformation indeed maps the CTC-free D1-D5-P superstratum onto a CTC-free solution in the M-theory frame.

\section{Comments on symmetric product orbifold CFTs}

It is a tantalizing prospect that the D1-D5-KKM system might have a solvable CFT in its moduli space, given that it is so similar to the D1-D5 system---differing only by a discrete identification on the transverse angular $\IS^3$.  One might think that since, in the decoupling limit, the introduction of KKM charge to the D1-D5 geometry amounts to a $\IZ_{n_k}$ orbifold of the Hopf fiber of $\IS^3$, that a similar quotient of the dual CFT by a chiral $\cR$-symmetry rotation would yield the corresponding dual CFT for the D1-D5-KKM system~\cite{Sugawara:1999qp}.\footnote{For related work on the microstates of the D1-D5-KKM system, see for example~\cite{Bena:2005ay,Saxena:2005uk,Balasubramanian:2006gi,Raeymaekers:2007ga,Giusto:2007tt,AlAlawi:2009qe,Banerjee:2014hza}.}

The first part of the construction in this paper maps a multi-wound D1-D5 supertube to a D1-D5-KKM bound state. It is tempting to translate this into a map between states of the D1-D5 symmetric product orbifold CFT and the putative D1-D5-KKM symmetric product orbifold CFT.
The multi-wound D1-D5 supertube configuration described in Section \ref{sec:supertube} corresponds to a R-R ground state of the D1-D5 CFT with $N_1 N_5/\kappa$ strands each of winding $\kappa$, with the same R-R ground state on each strand.

If there existed a symmetric product D1-D5-KKM CFT, for $n_1$ D1-branes and $n_5$ D5-branes and KKM charge $n_k = \kappa$, then this CFT should have total number of strands 
\bea
n_1 n_5 \kappa ~=~ \frac{N_1}{\kappa}\frac{N_5}{\kappa}\kappa
\eea
where we have used the relation between the brane numbers on the two sides of our map, given in equation~\eq{eq:KKMrescaledintegercharges}.
The map appears to conserve the total number of strands, while mapping strands of winding $\kappa$ in the D1-D5 CFT to strands of winding 1 in the D1-D5-KKM CFT. 

However, when $\cM=\IT^4$, strong constraints arise from the structure of U(1) currents and the energetics of states carrying the corresponding charges~\cite{Larsen:1999uk,Larsen:1999dh}, as we now review.

\subsection{Review of the D1-D5 CFT}

To begin, consider type IIB supergravity compactified on $\IT^5$.  The moduli space of this theory is the the 42-dimensional $\Gamma\backslash E_{6(6)}/USp(8)$, 
where the U-duality group $\Gamma$ is $E_{6(6)}(\IZ)$.  Wrapped branes and momentum excitations transform as a $\underline{27}$ under this group; the presence of the background D1-D5 charge vector $\vec q$ reduces the moduli space to the 20-dimensional $\cH_{\vec q}\backslash SO(5,4)/(SO(5)\!\times\! SO(4))$ through the attractor mechanism~\cite{Ferrara:1995ih}, and the U-duality group reduces to the subgroup $\cH_{\vec q}\subset SO(5,4;\IZ) \subset\Gamma$ that fixes $\vec q$.  The charge vector decomposes as 
\be
\underline{27}\rightarrow (\underline{1}\oplus \underline{9})\oplus \underline{16}\oplus \underline{1}
\ee
where the $\underline{1}\oplus \underline{9}$ represent the ``heavy" charges (branes wrapping the $y$ circle, including the D1-D5 background; the second singlet is the momentum charge along the $y$ circle; and the $\underline{16}$ comprises branes and momentum along the $\IT^4$ but {\it not} along the $y$ circle.

Elements of $SO(5,4;\IZ)$ not in $\cH_{\vec q}$ do not preserve the charge vector $\vec q$, instead they act as finite motions on the moduli space $\cH_{\vec q}\backslash SO(5,4)/(SO(5)\!\times\! SO(4))$.  Such transformations are {\it not} symmetries of the CFT, any more than any other finite motion on the moduli space preserves the CFT.  What such finite motions do tell us is that, if there is a weak-coupling cusp in the moduli space for a given pair of brane quanta $(n_1,n_5)$, then there are other cusps in the moduli space where the dual CFT becomes weakly coupled, one for each factorization of $N=N_1N_5$ into any other pair of integers $(N_1',N_5')$ with $N=N_1'N_5'$~\cite{Seiberg:1999xz,Larsen:1999uk}.  Because these are motions on the moduli space and not symmetries, the existence of a locus in the moduli space described by a symmetric product orbifold in one cusp does {\it not} imply the existence of such a description in any other cusp.

The question then arises, in which cusp does the symmetric product orbifold $(\IT^4)^N\!/S_N$ lie?  The BPS mass formula for the $\underline{27}$ is on one hand protected by supersymmetry, and on the other hand depends on the moduli and so determines the answer~\cite{Larsen:1999uk}.  In the decoupling limit of the D1-D5 system, the energetics of the $\underline{16}$ is (for a rectangular torus with all the antisymmetric tensor moduli switched off)
\be
h_R = \frac{1}{4 N_1}\sum_{i=1}^4 \Bigl(p_i\frac{\sqrt\gstr}{r_i} +  w^i_{D1} \frac{r_i}{\sqrt\gstr}\Bigr)^2
+ \frac{1}{4 N_5}\sum_{i=1}^4 \Bigl(w^i_{F1} \sqrt{\frac\gstr{v_4}}r_i +  w_i^{D3}\sqrt{\frac{v_4}\gstr}\frac1{r_i} \Bigr)^2  ~.
\label{eq:hR}
\ee
There is no invariant notion of the ``level'' of a U(1) current algebra, as the normalization of the current-current two-point function is moduli dependent.  For instance, from the previous considerations we know that $SO(5,4;\IZ)$ transformations can change the values of $n_1$ and $n_5$ in the above formula.  One can however compare the energetics of charged states in the CFT with the above expression.
The symmetric product orbifold has four left-moving translation currents (the diagonal sum of the translation currents in each copy of $\IT^4$), which realize the first of the two terms in Eq.\;\eq{eq:hR}, if we set $N_1=N$.
The other eight charges can be realized as the winding and momentum charges on a separate copy of $\IT^4$.  The presence of this additional component of the CFT is necessary to realize all the U(1) currents and the wrapped brane charges they couple to.
Thus it is natural to associate the symmetric product orbifold with the weak-coupling cusp of the moduli space where the appropriate low-energy description has $N_1=N$ and $N_5=1$. 

\subsection{The D1-D5-KKM CFT}

The addition of KK-monopole charge compactifies one more dimension of the target space -- the fibered circle of the KK-monopole (the $\psi$ circle), which is the Hopf fiber of $\IS^3$ in the decoupling limit.  One now has type IIB supergravity compactified on $\IT^6$, whose moduli space is $E_{7(7)}/SU(8)$.  The charge vector $\vec q$ of wrapped branes and momentum on $\IT^6$ transforms as a $\underline{56}$ of $E_{7(7)}$.  The background D1-D5-KKM charges break the moduli space down to the 28-dimensional space 
$\cH_{\vec q}\backslash F_{4(4)}/(SU(2)\!\times\! U\!Sp(6))$, 
and the $\underline{56}$ decomposes as
\be
\underline{56}\rightarrow (\underline{1} \oplus \underline{26}) \oplus (\underline{1} \oplus\underline{26}) \oplus \underline{1} \oplus \underline{1}
\ee
where once again the first $(\underline{1} \oplus \underline{26})$ is associated to the heavy background of branes wrapping the $y$ circle, and the second such factor is associated to wrapped branes and momentum along the compactification $\IS^1_\psi\times\IT^4$ transverse to the $y$ circle; the remaining two charges are KK-monopoles whose fibered circle is the $y$ circle, and momentum along the $y$ circle.  The $(\underline{1} \oplus \underline{26})$ of wrapped branes/momentum charges along $\IS^1_\psi\times\IT^4$ are again associated to a set of $U(1)$ currents in the CFT, and once again their energetics can be deduced from the decoupling limit of the BPS mass formula~\cite{Larsen:1999dh}
\begin{align}
\label{4chargemass}
h_R =& 
\frac{1}{4 n_1}\sum_{i=1}^4 \Bigl(p_i\frac{\sqrt\gstr}{r_i} +  w^i_{D1} \frac{r_i}{\sqrt\gstr}\Bigr)^2
+ \frac{1}{4 n_5}\sum_{i=1}^4 \Bigl(w^i_{F1} \sqrt{\frac\gstr{v_4}}r_i +  w_i^{D3}\sqrt{\frac{v_4}\gstr}\frac1{r_i} \Bigr)^2
\\
& \hskip 1cm + \frac{1}{4 n_k}\sum_{\sigma=1}^4 \Bigl(w^\sigma_{D} \, ne_\sigma +  w_\sigma^{\tilde D} \frac1{e_\sigma} \Bigr)^2
+ \frac{1}{12 n_1 n_5 n_k}\Bigl(d1_\psi \,n_5 + p_\psi \,n_k + d5_{\psi 6789} \,n_1 \Bigr)^2  ~.
\nn
\end{align}
Here the third octet of charges related to U(1)'s of ``level'' $n_k$ are $(f1_\psi, n5_{\psi 6789}, d3_{\psi ij})$, and the $e_\sigma$ are the corresponding volumes of the cycles they wrap, in appropriate units.

Once again there is a cusp of the moduli space for every factorization of $N$ into a triplet of background charges $n_1$, $n_5$, and $n_k$; the supergravity description of the CFT is thus merely a low-energy effective field theory approximation.  This fact also leads to a minor puzzle.  The only remnant of KK monopoles in the decoupling limit of the background is a $\IZ_{n_k}$ quotient of the angular $\IS^3$, which breaks the $SU(2)_L\!\times\! SU(2)_R$ $\cR$-symmetry down to $U(1)_L\!\times\! SU(2)_R$, and the supersymmetry from $(4,4)$ to $(0,4)$.   But when $n_k=1$, there is no quotient, and so it seems that there is an unbroken $(4,4)$ supersymmetry.  The resolution of this puzzle appears to be that indeed an accidental left-moving $\cN=4$ supersymmetry develops in the decoupling limit, on a codimension 8 sub-locus of the cusp of the moduli space corresponding to $n_k=1$.  The moduli space of the D1-D5-KKM system is 28-dimensional, in contrast to the 20-dimensional moduli space of the D1-D5 system; to get to any of the other supergravity limits with other values of $n_k$, one must turn on the additional eight moduli that break the accidental $\cN=4$ supersymmetry of the left-movers.

Again one can ask whether there is a symmetric product CFT ${\cW}^N/S_N$ somewhere in the moduli space.  It is again reasonable to suppose that the component CFT $\cW$ has four translation currents to generate the winding/momentum contributions in the first term of equation~\eqref{4chargemass}.  The diagonal current that survives the orbifold projection yields a U(1) of ``level'' $N$ and so can only match the above energetics in the cusp where one of the background charges is $N$, and again it is natural to take $n_5=n_k=1$ and $n_1=N$.  The second and third terms on the RHS are then the contributions of eight more currents of level one, and can be realized with a separate $\IT^8$ CFT.  

The last term in the wrapped brane energetics~\eqref{4chargemass} is difficult to realize in a symmetric product structure.  With $n_1=N$, $n_5=n_k=1$, one seeks another current of level $N$. 
If the building block is a $c=6$ superconformal field theory on $\IT^4$ (with once again an extra $\IT^4\times\IT^4$ CFT to realize the ``level-one'' terms), the translation currents comprise $c=4$, and their superpartners are four free fermions comprising the remaining $c=2$ (at least for the right-moving supersymmetric chirality).  Bilinears in the free fermions form a level-one SO(4)=SU(2)$\times$SU(2) current algebra, of which one SU(2) is the $\cR$-symmetry.  The other, ``auxiliary'' SU(2) has energetics $m^2/4$ for the individual component CFT $\cW$, where $m$ is the eigenvalue of $J_3^{\rm aux}$ for this auxiliary (level-one) SU(2) current algebra.%
\footnote{In the symmetric orbifold describing the D1-D5 system, this auxiliary SU(2) is an accidental symmetry of the orbifold locus, and does not survive perturbations away from this locus.}
The symmetric product structure then leads to an energetics $m^2/4N$ under the diagonal $J_3^{\rm aux}$.  This energetics of SU(2) level-one current algebra is thus incompatible with the last term in equation~\eqref{4chargemass} by a factor of $3$, and any attempt to engineer the requisite normalization naively leads to a breaking of the (0,4) supersymmetry.

One can ask whether this lattice of auxiliary SU(2) charges with energies $m^2/4N$ is a sublattice of some larger lattice of CFT zero modes, which also contains the values present in~\eqref{4chargemass}.  The possibilities are constrained by the full structure of U(1) charges in supergravity.  
The $(\underline{1} \oplus \underline{26})$ charges of wrapped branes/momentum on $\IS^1_\psi\times\IT^4$ decompose as 
\be
(\underline1,\underline1)\oplus (\underline2,\underline6)\oplus (\underline1,\underline{14})
\ee
under the local $SU(2)\!\times\! U\!Sp(6)$ symmetry of the moduli space of the D1-D5-KKM background.  The thirteen right-moving currents account for $(\underline1,\underline1)\oplus(\underline2,\underline6)$, with the singlet associated to the last term in~\eqref{4chargemass} and the second factor associated to the translation currents on the various copies of $\IT^4$; the remaining $(\underline1,\underline{14})$ are related to left-moving currents, for which there is less information due to the lack of supersymmetry in that chirality of the CFT.  A reasonable assumption is that twelve of the $\underline{14}$ are the left-moving counterparts of the first three terms in~\eqref{4chargemass} where one flips the relative sign of the ``winding'' and ``momentum'' contributions.  There are two more special currents whose energetics can then be determined from the local $SU(2)\!\times\! U\!Sp(6)$, leading to~\cite{Larsen:1999dh}
\begin{align}
\label{leftmovers}
h_L =& 
\frac{1}{4 n_1}\sum_{i=1}^4 \Bigl(p_i\frac{\sqrt\gstr}{r_i} -  w^i_{D1} \frac{r_i}{\sqrt\gstr}\Bigr)^2
+ \frac{1}{4 n_5}\sum_{i=1}^4 \Bigl(w^i_{F1} \sqrt{\frac\gstr{v_4}}r_i -  w_i^{D3}\sqrt{\frac{v_4}\gstr}\frac1{r_i} \Bigr)^2
\\
& \hskip 1cm + \frac{1}{4 n_k}\sum_{\sigma=1}^4 \Bigl(w^\sigma_{D} \,e_\sigma -  w_\sigma^{\tilde D} \frac1{e_\sigma} \Bigr)^2 
+ \frac{1}{4 n_1 n_5 n_k} \Bigl(d1_\psi \, n_5 - p_\psi \, n_k\Bigr)^2
\nn\\
& \hskip 2cm + \frac{1}{12 n_1 n_5 n_k}\Bigl(d1_\psi \,n_5 + p_\psi \,n_k - 2 d5_{\psi 6789} \,n_1 \Bigr)^2  ~.
\nn
\end{align}

The spectrum of the one right-moving and two left-moving ``special'' currents associated to the charges $p_\psi,d1_\psi , d5_{\psi 6789}$ in~\eqref{4chargemass}, \eqref{leftmovers} has also arisen in a related context, in which spectral flows were used to generate a class of nonsupersymmetric solutions.%
\footnote{In comparing equation~\eqref{4chargemass} above to the spectrum equation~5.24 of~\cite{Banerjee:2014hza}, one notes a typo of a missing factor of 1/2 in the first term on the RHS of the latter.}
Indeed, the charged states are all non-BPS, even though the starting point in the analysis is a BPS mass formula; after the decoupling limit, none of the U(1) currents lie in the stress tensor supermultiplet, even though before the decoupling limit, the right-moving charges did have that property.  The U(1) charges in the CFT are thus no longer $\cR$-charges, and therefore there is no BPS condition involving them.  Spectral flow remains a robust property of the CFT that follows from symmetry, and leads to the same result as the combination of the decoupling limit of the BPS formula for the right-movers and the moduli space considerations employed in~\cite{Larsen:1999dh} to obtain the charge spectrum.  This gives us further confidence in the applicability of these formulae, though with the caveat that the full energy of any given state will  typically not be saturated by the contributions of the U(1) charges.

A $\IT^4$ symmetric product accounts for the first term in~\eqref{leftmovers} via the left-moving $\IT^4$ translation currents, and similarly the second and third terms correlate with the corresponding terms in~\eqref{4chargemass}.  This leaves two additional left-moving currents of level $N$.  The three ``special'' currents not associated to torus translations (two left-moving and one right-moving), plus the right-moving $\cR$-symmetry current, thus all have level $N$ and soak up all the central charge of the right-moving fermions in the symmetric product, and the corresponding remaining central charge of the left-movers.  Bosonizing all four currents leads to a (2,2) lattice of zero modes whose energetics must match~\eqref{4chargemass}, \eqref{leftmovers}.  

The energies of a general (2,2) lattice of zero-modes has the form
\begin{align}
\hlat_L &= \frac{1}{4\rho_2\tau_2} \bigl|(n_1 - \tau n_2)-\rho(m_2+\tau m_1)\bigr|^2
\nn\\
\hlat_R &= \frac{1}{4\rho_2\tau_2} \bigl|(n_1 - \tau n_2)-\bar\rho(m_2+\tau m_1)\bigr|^2
\end{align}
for complex $\rho=\rho_1+i\rho_2$, $\tau=\tau_1+i\tau_2$.  Without loss of generality, we can write 
\be
m_2 = \frac12(m_L+m_R) ~~, ~~~~ n_2=\frac12(m_L-m_R)
\ee
and interpret $m_R$ as the eigenvalue of $J^3_R$ of the $\cR$-symmetry.  Demanding that $m_R$ appear only in $h_R$ and only quadratically implies $\rho=\tau$.  The right-moving energetics~\eqref{4chargemass} is reproduced for $\tau_1=\rho_1=1$, $\tau_2=\rho_2=\sqrt{3}$
\begin{align}
\hlat_R &= \frac{m_R^2}{4} + \frac{(m_L+4m_1-n_1)^2}{12}
\end{align}
if we identify
\be
m_L+4m_1-n_1 = d1_\psi  + p_\psi + d5_{\psi 6789} N ~~.
\ee
Examining the contribution of the charges to the left- and right-moving energies, the closest match comes if we identify
\be
m_L = p_5-f_5 ~~,~~~~ m_L-n_1 = p_5+f_5 ~~,~~~~ 4m_1 = d5_{\psi 6789} N
\ee
which leads to a match between the lattice and supergravity expressions for the right-moving energy.  The difference between the supergravity and symmetric product formulae then becomes
\be
h_L - \hlat_L  = \frac{(N d5_{\psi 6789})^2}{4} - \frac{(N d5_{\psi 6789})p_5}{2}
\ee
which is reminiscent of the structure of a spectral flow. 
The low-lying spectrum (energies much less than order $N$) is only compatible with $d5_{\psi 6789}=0$.  The lattice of such states, when chosen to match the results of the BPS mass formula, cannot simultaneously accommodate the spectrum of free fermion superpartners of the torus translation currents.

To summarize, supersymmetry and a symmetric product of $c=6$ building blocks leads to a lattice of U(1) charges which is not compatible with the lattice inferred from supergravity considerations.
The right-moving fermions which are the superpartners of right-moving translation currents have $\cR$-charge 1/2 and dimension 1/2; on the other hand, that lattice of charges inferred from supergravity does not have such a state in its spectrum.  This throws considerable doubt on the existence of a symmetric product orbifold locus in the moduli space of the MSW CFT.

\newpage
\section{Discussion}
\label{sec:Disc}

Understanding the dynamics of multiple  M5 branes has been one of the most challenging and  interesting issues in string theory for quite a number of years.  There has been a huge effort in understanding how M5-brane theories can describe strongly-coupled gauge theories in four dimensions.  Our purpose in this paper has been to study what should, perhaps, be one of the simplest avatars of the M5-brane field theory:  The (1+1)-dimensional MSW CFT that comes from wrappings of an M5 brane on a very ample divisor of a Calabi-Yau manifold. This seemingly simple CFT remains enigmatic, almost twenty years after it was first shown to be able to encode microstate structure of four-dimensional black holes \cite{Maldacena:1997de}.  In this paper we have considered M5 branes wrapping $4$-cycles in $\IT^6$ and $\IT^2 \times $K3, but our resulting M-theory solutions can be trivially extended to compactifications with a more general field content.

As we have discussed, part of the difficulty in analyzing this CFT is that it does not seem to have any point in its moduli space with a canonical description in terms of better-understood conformal field theories, such as a symmetric orbifold theory.  However, one can use holographic methods to study this theory at strong coupling, and in this paper we have made significant progress in that direction:  We have obtained explicit families of smooth, horizonless solutions of five-dimensional supergravity that are dual to families of BPS states of the MSW CFT.

We constructed these families of solutions to M-theory by deriving a map between them and a class of states of the D1-D5 CFT, described as smooth, horizonless solutions to six-dimensional supergravity.  This was done by transforming asymptotically AdS$_3 \times \IS^3\times \IT^4$, D1-D5-P superstratum solutions that are independent of the Hopf fiber of the $\IS^3$ to asymptotically  AdS$_3 \times \IS^2\times \IT^6$ solutions\footnote{Our solutions can trivially be extended by replacing $\IT^4$ by K3 and $\IT^6$ by $\IT^2 \times $K3.} dual to momentum-carrying microstates of the MSW CFT.   We therefore referred to our new families of solutions as {\em M-theory superstrata}. In principle, one should be able to obtain families of M-theory superstrata that depend on arbitrary functions of two variables (with arbitrary Fourier modes around the axis of the $\IS^2$ and the spatial axis of the AdS$_3$).  In this paper we have constructed solutions which have single Fourier mode excitations.  However, based upon the success of the superstratum program in six dimensions \cite{Bena:2015bea}, we anticipate that one should be able to find smooth,  horizonless M-theory superstrata with general families of Fourier modes excited. 

It is important to emphasize that there are many more M-theory superstrata solutions constructed using our technology than those that we have directly mapped to smooth D1-D5-P superstrata. As we have seen in Section \ref{sec:GeneralContracting}, when the KKM charge, $\kappa$, is greater than one,  the smooth D1-D5-P superstrata map to M-theory superstrata with mode numbers along the AdS$_3$ circle that are multiples of $\kappa$.  However, once in the M-theory frame, nothing prevents us from extrapolating these solutions to generic values of the mode numbers compatible with smoothness and appropriate M-theory periodicities. Under our map these more generic M-theory superstrata do not transform into geometric D1-D5-P states\footnote{A naive application of our map would give rise to solutions with multivalued fields, and if one extrapolates the candidate dual CFT states of~\cite{Bena:2016agb,Bena:2016ypk} to the appropriate values of the parameters, one would not satisfy the condition of integer momentum per strand. Thus a straightforward application of this holographic dictionary suggests that these configurations should be discarded. For more discussion, see~\cite{Bena:2016agb}.}, and yet they are perfectly good solutions.\footnote{Rather than using our map, one could also obtain these solutions by setting $\kappa=1$ in the D1-D5 superstrata, restricting to $k=2m$, introducing the smooth $Z_\kappa$ quotient of the Hopf fiber by hand and U-dualizing.}

As we have noted in the Introduction, there have been several earlier approaches to the construction of solutions dual to momentum-carrying BPS microstates of the MSW CFT.  The common goal of this paper and of previous work has been to examine the spacetime structure of the microstates of black holes with a macroscopically-large horizon area.
In this system, these black holes have a momentum charge along the AdS$_3$ circle that, for given M5 charges, must be larger than a certain threshold, which is of order the product of the three M5 charges; once above this threshold, one is in the ``black hole regime'' of parameters.  
In Type IIA, the M5 and momentum charges become D4 and D0 charges. 

The microstate geometry corresponding to the maximally-spinning Ramond ground state of the MSW CFT is obtained by blowing up the single-center D4-D4-D4 configuration to a two-center fluxed D6-$\overline{\rm D6}$ configuration, whose M-theory uplift is [global AdS$_3] \times \IS^2 \times \IT^6$. The addition of D0 charge via back-reacted singular D0's was studied in~\cite{Denef:2007yt,deBoer:2008fk}.
The degeneracy of such ``D0-halo'' solutions was counted in~\cite{deBoer:2008zn,deBoer:2009un}, and found to give rise to an entropy that matches that of an M-theory supergraviton gas in [global AdS$_3] \times \IS^2 \times \IT^6$ for sufficiently small D0 charge.  The full back-reaction of the supergraviton gas states has never been computed, but since our M-theory superstratum solutions represent smooth waves in AdS$_3\times \IS^2$, one may expect that at least some of them can be thought of as coming from back-reacted supergraviton gas states. Furthermore, if the full non-Abelian and nonperturbative interactions of uplifted D0-branes results in solutions that are non-singular and varying along the M-theory circle, one expects these solutions to also resemble our M-theory superstrata. Hence, it may be that the smooth back-reacted solutions we construct are the missing link needed to connect the entropy counts in the (non-back-reacted) supergraviton gas and (singular) D0-halo approaches. 

In the black-hole regime of parameters, the D0-halo entropy exhibits a sub-leading growth with the charges compared to the black hole entropy~\cite{deBoer:2008fk}.  On the other hand, in this regime the solutions have large deep AdS$_2$ throats with high redshifts, and so are no longer small perturbations of [global AdS$_3] \times \IS^2 \times \IT^6$. A robust estimate of the number of states comprised by superstrata remains to be carried out.

One can also add momentum by adding M2-branes that wrap the two-sphere of the AdS$_3\times \IS^2$, and that carry angular momentum on both the AdS$_3$ and the $\IS^2$~\cite{Denef:2007yt}.  The entropy of these configurations comes from the high degeneracy of the Landau levels that result from the dynamics of the M2-branes on the compactification manifold in the presence of M5 flux~\cite{Gaiotto:2004ij,Denef:2007yt,Martinec:2015pfa}, and has been argued to scale in the same way as that of the black hole. The back-reaction of these ``W-brane'' configurations is fully worked out only in some very simple examples \cite{Raeymaekers:2015sba}. However, on general grounds one expects uncancelled tadpoles which give rise to asymptotics that are different from the asymptotics of bulk duals of MSW CFT states.  If, on the other hand, one cancels the tadpoles using additional brane sources, there are no preserved supersymmetries whatsoever~\cite{Tyukov:2016cbz}. Furthermore, in more generic multi-center solutions, the corresponding W-brane configurations also give rise to tadpoles, which can only cancel when the W-branes form a closed path among the centers.\footnote{The counting of these closed paths gives an entropy that scales in the same way as the black hole entropy as a function of the charges~\cite{Denef:2007vg,Bena:2012hf,Martinec:2015pfa}.} Hence, when the multi-center solution has a throat of finite length, these additional M2-brane bound states break at least another half of supersymmetry (giving \nBPS{16} states), and typically all of the supersymmetry~\cite{Tyukov:2016cbz}. Thus these states cannot correspond to microstates of the BPS MSW black hole. 

Given the large entropy of the  W-branes, one would like to somehow restore the broken supersymmetry. This can only be achieved by going to a scaling limit, in which the throat becomes infinitely deep.  In the infinite-throat limit, the solitonic W-branes become massless, new dynamical fields emerge (corresponding to the Higgs branch of the field theory for which the W-branes are individual quanta) and the rich families of W-branes become reflections of the rich degeneracies of the vacua of these new dynamical fields.  Thus W-branes should provide a semi-classical way to access the Higgs branch~\cite{Martinec:2015pfa}.

Another way to access the physics of the Higgs branch is via world-sheet disk amplitudes. Using these techniques one can compute the supergravity back-reaction of D-brane bound states upon an infinitesimal displacement on the Higgs branch.
In the D1-D5 system, such calculations demonstrate that the additional tensor multiplet described by $(Z_4, G_4)$ is an integral part of the back-reaction of generic Higgs-branch states \cite{Giusto:2011fy}. Thus one expects the configurations that result from condensing the W-branes to include such additional species of supergravity fields.
For  four-charge black holes in four dimensions, in the D3-D3-D3-D3 system (which is U-dual to the D1-D5-KKM-P and the M5-M5-M5-P systems), a similar string emission calculation was recently performed~\cite{Bianchi:2016bgx,Bianchi:private}, confirming the presence of this kind of additional species of supergravity fields in the backreaction of these bound states.
	
Remarkably, these new species of supergravity fields are exactly those needed to give rise to smooth superstrata solutions, via the coiffuring procedure we have used in Sections \ref{sec:GeneralContracting} and \ref{sec:example}. Furthermore, if one back-reacts M5 branes of~\cite{Maldacena:1997de} wrapping smooth ample divisors inside $\IT^6$, one expects to source exactly these additional supergravity fields. 
If one combines these two features with the fact that our M-theory superstrata solutions should be parameterized by arbitrary continuous functions, and hence have a large entropy, it appears very likely that these supergravity fields are a key component of the structure of typical black hole microstates.

Our results raise some interesting questions about the formal mathematical structures of five-dimensional supergravity solutions.  One should recall that the construction of smooth microstate geometries in five dimensions was done via locally hyper-K\"ahler base metrics whose signature changes from $+4$ to $-4$ on certain hypersurfaces. These singular base metrics are referred to as ambi-polar or pseudo-hyper-K\"ahler base metrics, and the hypersurfaces where the signature changes are referred to as ``degeneration loci''. 
While the four-dimensional spatial base metric is singular, all singularities cancel in the five-dimensional Lorentzian metric. 
There has been a growing mathematical interest in the geometry of  these ambi-polar spaces~\cite{Niehoff:2016gbi}, generalizing the notion of ``folded'' hyper-K\"ahler metrics~\cite{Hitchin1,Biquard1}.  Our results here indicate that harmonic analysis on such manifolds might be extremely rich and interesting. 

In particular, the first step in solving the BPS equations is to find smooth, harmonic two-forms on the spatial base metric.  In standard Riemannian geometry, this is a classical exercise and the harmonic forms are dual to the homology cycles.   The original work on microstate geometries involving ambi-polar bases \cite{Bena:2005va,Berglund:2005vb,Bena:2007kg} simply translated the expressions for the standard harmonic forms of Riemannian geometry.
The solutions constructed in this paper have only one homology cycle, but we have exhibited infinitely many ``pseudo-harmonic'' two-forms.  We defined such two-forms to be those that are closed and co-closed (``harmonic''), potentially singular on the degeneration loci of the base geometry, and yet lead to  completely regular, five-dimensional BPS solutions.   This leads to several interesting questions.  Firstly, how does our result generalize to multi-centered ambi-polar GH metrics?   More generally, what is the classification of  pseudo-harmonic two-forms?  This paper shows that what seems to be a rigid topological problem actually has an infinite amount of ``wiggle room'' on an ambi-polar base.

Returning to our map between states of the MSW and D1-D5 CFT's, the results presented here suggest that this map should contribute more deeply to our understanding of the physics of four-charge black holes in four dimensions and to the question of how much entropy of these black holes comes from smooth horizonless solutions.   More broadly, we believe that our map will also prove useful in gaining deeper understanding of the hitherto mysterious MSW CFT.  As we have seen, only a particular class of the MSW microstate geometries are related to D1-D5-P ones, and hence only a sub-sector of the states of the MSW CFT is mapped to a sub-sector of the D1-D5 CFT.  It would be extremely interesting to explore and test possible extensions of this correspondence. 

Indeed, several important questions remain about our map. First, the map is defined in terms of geometrical data, and it is interesting to see whether one can generalize it to other CFT states that are not dual to smooth torus-independent horizonless supergravity solutions, but may involve string or brane degrees of freedom, dependence on the internal directions, or high-curvature corrections\footnote{For example, one can imagine constructing ten-dimensional supergravity solutions dual to D1-D5 microstates that have a non-trivial dependence on the torus coordinates, and therefore cannot be described in a six-dimensional truncation.  Our map would take these solutions into holographic duals of MSW microstates that contain an infinite tower of Kaluza-Klein modes, and thus cannot be described in five-dimensional supergravity.}. 
A pessimistic possibility is that our construction is merely an approximation that relates particular geometrical solutions in the supergravity limit, but that does not map CFT physics beyond small perturbations around the particular states that can be related to each other. 
On the other hand, it is tempting to speculate that, if one accepts holography as a correct description of all physics in asymptotically-AdS backgrounds (including all $1/N$ corrections), a generalization of our map to degrees of freedom beyond six-dimensional supergravity may exist, and it would be interesting to investigate its properties. 

A related question is whether our map is simply a useful device for counting and classifying certain MSW states, or whether it is capable of capturing other CFT data such as anomalous dimensions or three-point functions.  
In the supergravity approximation, these quantities can in principle be computed perturbatively around a given solution~\cite{Witten:1998qj}, giving one hope that additional information about the MSW CFT could be gleaned.

One can reasonably expect at least some three-point functions to be mapped from one sector of one CFT to another sector of the other CFT, because of non-renormalization theorems~\cite{Baggio:2012rr}. However, if one considers the four-point functions of an MSW operator that gets mapped to a D1-D5 one, these four-point functions are computed by summing over all operators in the intermediate channel, which may not belong to the relevant sub-sectors. Furthermore, generic four-point functions are not protected when one deforms away from the free orbifold point of the D1-D5 CFT to the supergravity point, and hence there is no reason to expect a map for this data. Nevertheless, one might hope to use our map to find a prescription that allows one to calculate at least certain conformal blocks of the MSW CFT from D1-D5 ones, which would already be remarkable progress.

What is clear is that, as a CFT-to-CFT map, our construction is quite unusual. Indeed, to go from the D1-D5 NS vacuum to the MSW one, one needs to perform a combination of spectral flow transformations and ``gauge'' transformations. While spectral flow transformations have a clear CFT interpretation, as the redefining of the CFT Hamiltonian by the addition of a term proportional to the $\cR$-charge, the gauge transformation would appear to correspond to redefining the $\cR$-charge by the addition of a term proportional to the Hamiltonian, which is much more mysterious.  Hence, while spectral flow is an operation that maps states to states within the CFT, the gauge transformation appears to change the CFT itself.  On the other hand, since the MSW CFT does not appear to have any point in its moduli space with a symmetric product orbifold description, a map of the type we have found may be the most one can hope for.

\section*{Acknowledgments}

\vspace{-2mm}
We thank Massimo Bianchi, Duiliu Emanuel Diaconescu, Stefano Giusto, Monica Guic\u{a}, Stefanos Katmadas, David Kutasov, Jos\`e Francisco Morales, Rodolfo Russo, Masaki Shigemori and Amitabh Virmani for valuable discussions. 
EM and NPW are very grateful to the IPhT, CEA-Saclay for hospitality during the initial stages of this project.
The work of IB and DT was supported by the John Templeton Foundation Grant 48222 and by the
ANR grant Black-dS-String. 
The work of NPW was supported in part by the DOE grant DE-SC0011687;
that of EJM was supported in part by DOE grant DE-SC0009924.
The work of DT was further supported by a  CEA Enhanced Eurotalents Fellowship.

\newpage
\begin{appendix}

\section{Covariant form of BPS ansatz and equations}
\label{app:covariant}

To rewrite our ansatz in covariant form, we rescale $(Z_4,\Theta_4,G_4) \to (Z_4,\Theta_4,G_4)/\sqrt{2}$. 
Then we have 
\be
C_{123} ~=~ 1 \,, \qquad C_{344} ~=~ -1 \,. 
\ee
It should be understood that this rescaling holds throughout this Appendix (and only in this Appendix).
Then we define the (mostly-minus, light-cone) $SO(1,2)$ Minkowski metric via
\bea
\eta_{ab} &=& C_{3ab} \qquad \Rightarrow \qquad \eta_{12} ~=~ \eta_{21} ~=~ 1\,, \quad \eta_{33} = -1 \,.
\eea
which can be used to raise and lower $a,b$ indices, now that the above rescaling has been done. After the rescaling we have
\be
\cP ~=~ \coeff{1}{2} \eta^{ab} Z_a Z_b ~=~ Z_1 Z_2 - \coeff12 Z_4^2 \,.
\ee
The first layer of the BPS equations then takes the form
\bea
 *_4 D\dot{Z}_a ~=~ & \eta_{ab} D\Theta^{(b)}\,,\qquad D*_4 DZ_a ~=~  - \eta_{ab} \Theta^{(b)} \! \wedge d\beta\,,
\qquad \Theta^{(a)} ~=~ *_4 \Theta^{(a)} \,.
\eea
The second layer becomes
\begin{equation}
 \begin{aligned}
D \omega + *_4 D\omega + \mathcal{F} \,d\beta 
~=~ & Z_a \Theta^{(a)}\,,  \\ 
 *_4D*_4\!\Bigl(\dot{\omega} -\coeff{1}{2}\,D\mathcal{F}\Bigr) 
~=~& \ddot \cP  -(\dot{Z}_1\dot{Z}_2  - \coeff{1}{2}(\dot{Z}_4)^2 )
-\coeff{1}{4} \eta_{ab} *_4\! \Theta^{(a)}\wedge \Theta^{(b)} \,.
\end{aligned}
\label{eqFomega-app}
\end{equation} 
Our ansatz for the tensor fields is
\bea
G^{(a)}  &=&  d \left[ - \frac{1}{2}\,\frac{\eta^{ab} Z_b}{\cal P}\,(du + \omega ) \wedge (dv + \beta)\, \right] ~+~\coeff{1}{2} \;\! \eta^{ab} *_4 \! D Z_b  
~+~ \coeff{1}{2}\,  (dv+ \beta) \wedge \Theta^{(a)} \,.
\label{G-ans-cov}
\eea
In our conventions, the twisted self-duality condition for the field strengths is 
\be
\ast_6 G^{(a)} ~=~ \T{M}{a}{b} G^{(b)} \,, \qquad\qquad  M_{ab} ~=~ \frac{Z_a Z_b}{\cP} - \eta_{ab} \,.
\ee

\section{Circular D1-D5 supertube: parameterizations}
\label{app:supertube}

In this appendix we recall the usual representation of a circular D1-D5 supertube solution within the six-dimensional ansatz \eq{sixmet}, and the relation to the representation used in this paper. 

The usual representation (see, for example,~\cite{Bena:2015bea,Bena:2016agb}) is given using the coordinate transformation (\ref{tyuv-1}) and setting 
 $\mathcal{F} = 0$. This solution is then given by:
\begin{eqnarray}
Z_1  &=& \frac{Q_1}{\Sigma} \,, \qquad Z_2  ~=~ \frac{Q_5}{\Sigma}\,, \qquad 
\cF ~=~ 0 \,, \qquad Z_4  ~=~0 \,;  \qquad\quad \Theta^{(j)} ~=~ 0\,, \quad j = 1,2,4  \,;
\cr
\beta &=& \frac{ \kappa R_y a^2 }{\sqrt{2} \;\! \Sigma} \, ( \sin^2 \theta \, d\varphi_1-   \cos^2 \theta \, d\varphi_2 )\,, \qquad  
\omega  ~=~\frac{\kappa R_y a^2 }{\sqrt{2} \;\! \Sigma} \, ( \sin^2 \theta \, d\varphi_1 +   \cos^2 \theta \, d\varphi_2 ) \,. 
\label{eq:betaomegaST-app}
\end{eqnarray}
Using the transformations (\ref{shiftinv-app}) and (\ref{rescalings}), we obtain the following solution:
\begin{eqnarray}
\tilde{u} &=& t \,, \qquad \tilde{v} ~=~ t+y \,,  \cr
Z_1  &=& \frac{Q_1}{\Sigma} \,, \qquad Z_2  ~=~ \frac{Q_5}{\Sigma}\,, \qquad 
\medtilde{\cF} ~=~ -1 \,, \qquad Z_4  ~=~0 \,;  \qquad\quad \Theta^{(j)} ~=~ 0\,, \quad j = 1,2,4  \,;
\cr
\medtilde{\beta} &=& \frac{ \kappa R_y a^2 }{\Sigma} \, ( \sin^2 \theta \, d\varphi_1-   \cos^2 \theta \, d\varphi_2 )\,, \qquad  
\medtilde{\omega} = \frac{\omega + \beta}{\sqrt{2}}  ~=~\frac{\kappa R_y a^2 }{\Sigma} \,\sin^2 \theta \, d\varphi_1 \,.  
\label{eq:betaomegaST-app-2}
\end{eqnarray}
This is the form of the solution used in the main text in \eq{eq:betaomegaST}.

For superstratum solutions, the same redefinitions can be applied, and then the fields asymptote to the form \eq{eq:betaomegaST-app-2}.

\section{Lattice of identifications and fractional spectral flow}
\label{app:lattice-spectral}

In this appendix we illustrate the step of redefining the lattice of identifications, with the explicit example of fractional spectral flow of a multi-wound circular D1-D5 supertube solution~\cite{Giusto:2012yz}.

The starting configuration is the multi-wound circular D1-D5 supertube in the decoupling limit, given in Eqs.\;\eq{eq:ST-harm-fns}--\eq{streg}. To this solution we apply a (fractional) spectral flow transformation \eq{eq:SF-harm} with parameters $(\gamma_1,\gamma_2,\gamma_3) = (0,0,-s/(\kappa R))$, together with an accompanying gauge transformation. The details and the resulting harmonic functions can be found in Appendix A of~\cite{Giusto:2012yz}. Recall that we have defined $R\equiv R_y/2$.

The point that we emphasize here is that to generate the transformed solution, one inserts these new harmonic functions into a ``hatted'' version of the general ansatz, as in \eq{recastsixmet}, and importantly, one takes the lattice of identifications to be the standard one in the hatted coordinates:
\be \label{eq:new-lattice}
\hat{v} ~\sim~ \hat{v} + 2 \pi R_y \,, \qquad \hat{\psi} ~\sim~ \hat{\psi} + 4\pi \,.
\ee
Equivalently, one can use the coordinate form of this fractional spectral flow transformation, namely
\be
\psi ~=~ \hat{\psi} - \frac{s}{\kappa R}\hat{v}  \,, \qquad\quad v ~=~ \hat{v} \,.
\ee
and act with it on the explicit multi-wound circular supertube metric~\eq{AdStimesS-2b}, again imposing \eq{eq:new-lattice}.

Since $\hat{\psi}$ is the only coordinate that has transformed non-trivially, for ease of notation we shall re-use the coordinates of the starting solution $\zeta$, $\eta$, $\tilde{t}$, $\tilde{y}$ defined in \eq{eq:tildecoords}, as well as $\phi$, without writing hats explicitly. Then the transformed decoupling-limit solution is:
\begin{eqnarray}
ds_6^2  &\!=\!&  \sqrt{Q_1 Q_5}\left[ -  \cosh^2 \zeta \, {d\tilde{t}}^2 + d \zeta^2
 +   \sinh^2 \zeta \,  d\tilde{y}^2 \right] 
\label{AdStimesS-s}\\
&& \hspace{-8mm}
{} + \!\frac{\sqrt{Q_1 Q_5}}{4} \! \left[ \!
\Big( \! \big[d\hat\psi - (2s+1)(d\tilde{t}+d\tilde{y}) \big] \!\!\;+ \cos \eta  \!\!\; \big[ d\phi +(d\tilde{t}-d\tilde{y}) \big] \Big)^2  
\!\!\!\; +  d\eta^2 \!\!\; + \sin^2 \eta \big[ d\phi +(d\tilde{t}-d\tilde{y}) \big]^2  \!\!\;
  \right] .
\nonumber
\end{eqnarray}
As in the starting solution ($s=0$), there is a coordinate change to bring the metric to local AdS$_3 \times \IS^3$ form. For the above transformed solution, it is of course
\be
\psi' \,=\, \hat{\psi} - (2s+1)(\tilde{t}+\tilde{y}) \,, \qquad\quad  \phi' \,=\, \phi + (\tilde{t}-\tilde{y})\,.
\label{eq:map-s-to-local-AdS3}
\ee
The combination of \eq{eq:new-lattice} and \eq{eq:map-s-to-local-AdS3} gives rise to an interesting variety of orbifold singularities in the core of these solutions, depending on the common divisors of the integer parameters $s$, $s+1$ and $\kappa$, as noted in~\cite{Jejjala:2005yu} and analyzed in detail in~\cite{Giusto:2012yz}.

\section{MSW maximal-charge Ramond ground state solution} 
\label{app:MSW-vac}

\subsectionmod{Coordinate conventions}

We record here for convenience some of our coordinate conventions.
We define the three-dimensional distances $r_{\pm}$ from the centers, in Cartesian and cylindrical coordinates:
\begin{equation}
r_\pm ~\,\equiv\,~  \sqrt{y_1^2 + y_2^2 + (y_3 \mp c)^2} ~\,\equiv\,~  \sqrt{\rho^2 \,+\, (z\mp c)^2}\,, \qquad\quad c \;=\; \coeff{1}{8} \, a^2 \,. \label{rpmdefn}
\end{equation}
We define angular coordinates measured from the $z=y_3$ axis at the two centers via
\be
\cos \vartheta_{\pm} ~=~ \frac{z\mp c}{r_{\pm}} \,.
\ee
The relation between these coordinates and the $(r,\theta)$ coordinates used throughout the paper is
\be
\Sigma ~\equiv~ 4 r_-  ~\equiv~   (r^2 + a^2 \cos^2 \theta) \,, \qquad   \Lambda ~\equiv~4 r_+ ~\equiv~(r^2 + a^2 \sin^2 \theta)  \,, \label{LamSigdefn-app}
\ee
\vspace{-9mm}
\begin{eqnarray}
\cos \coeff{1}{2}  \vartheta_{-} &=& \frac{(r^2 + a^2)^{1/2}}{\Sigma^{1/2}}\, \cos \theta  
\,, \qquad\quad~  
\sin \coeff{1}{2} \vartheta_{-} ~=~ \frac{r}{\Sigma^{1/2}}\, \sin \theta 
\,,  \cr
\cos \coeff{1}{2}  \vartheta_{+} &=& \frac{r}{\Lambda^{1/2}}\, \cos \theta  
\,, \qquad\qquad\qquad~  
\sin \coeff{1}{2} \vartheta_{+} ~=~ \frac{(r^2 + a^2)^{1/2}}{\Lambda^{1/2}}\, \sin \theta 
\,. \label{coords1-app}%
\end{eqnarray}

Prolate spheroidal coordinates centered on $r_\pm = 0$ are useful for writing the metric as global AdS$_3$:
\begin{equation}
 z =  c\, \cosh 2\zeta \,\cos \eta \,, \qquad  \rho =  c\, \sinh 2\zeta \, \sin \eta \,, \qquad
 \zeta \ge 0\,, \ \ 0 \le \eta \le \pi \,.
  \label{coordsa}
 \end{equation}
In particular, one has 
\begin{equation}
r_\pm   ~=~  c\, (\cosh 2\zeta  \mp \cos \eta) \,.
  \label{rpmform}
 \end{equation}

\subsectionmod{MSW maximal-charge Ramond ground state solution}

The five-dimensional MSW maximal-charge Ramond ground state solution is described by the following harmonic functions.
Using $I=1,2,3$ and employing notation mod 3 for the $I$ indices, we have
\begin{eqnarray}
 V &=&  \frac{1}{r_+} - \frac{1}{r_-} \,, \qquad\qquad\qquad\qquad~\,   
K^I ~=~ \frac{k^I}{2} \left( \frac{1}{r_+} +  \frac{1}{r_-} \right) \,,  \cr
 L_I &=& -\frac{k^{I+1} k^{I+2}}{4}  \left( \frac{1}{r_+} - \frac{1}{r_-} \right) \,,
 \qquad 
M ~=~  \frac{k^1k^2k^3}{8} \left( \frac{1}{r_+} +  \frac{1}{r_-} \right)  - \frac{k^1k^2k^3}{2c} \,.
\label{eq:functdec-app}
\end{eqnarray}
The four-dimensional base metric can be written as
\begin{equation}
ds_4^2 ~=~ V^{-1} \, \big( d\psi + A)^2  ~+~ V\, (d\rho^2 + dz^2 + \rho^2 d \phi^2) \,. \label{GHpolar}
\end{equation}
We write the one-form $A$ as 
\be
A ~=~ (\cos \vartheta_+ - \cos\vartheta_- )d\phi \,.
\ee
Note that in this gauge, near the GH centers, $A \simeq (-1 \pm \cos\vartheta_{\pm})d\phi$, so the lattice of identifications that gives smoothness is (c.f.~Footnote\;\ref{foot:psi-phi-lattice})
\be
\psi ~\cong~\psi+4\pi \,, \qquad\quad \phi ~\cong~ \phi+2\pi \,.
\label{eq:psi-phi-lattice-MSW-app}
\ee
The one-form $\varpi$ is given by
\begin{equation}
\varpi ~=~   - \frac{k^1k^2k^3}{4c} \left( \frac{\rho^2 + (z-c +r_+)  (z+c - r_-)}{r_+ r_-}  \right) d \phi\,. 
\end{equation}
The metric of this solution is that of global AdS$_3 \times\IS^2$. To see this, we pass to the prolate spheroidal coordinates $(\zeta,\eta)$ defined in \eq{coordsa}, and define the coordinates
\begin{equation}
 \tau ~\equiv~   \frac{c}{k^1k^2k^3}t  \,, \qquad 
\varphi ~\equiv~   \frac12 \psi -  \tau    \,, \qquad 
	\tilde{\phi} ~\equiv~ \phi + \tau -  \varphi  \,,
 \label{coordsb}
 \end{equation}
in terms of which the five-dimensional metric is manifestly global AdS$_3 \times\IS^2$,
\begin{equation}
ds_5^2 ~=~ R_1^2 \Big( - \cosh^2\zeta \,  d\tau^2 + d\zeta^2 +  \sinh^2 \zeta \, d\varphi^2 \Big) ~+~  R_2^2 \left(   d \eta ^2 + \sin^2\eta  \, d\tilde\phi^2 \right)  \,,
 \label{AdS3S2appendix}
 \end{equation}
with
\begin{equation}
  R_1~=~  2 R_2 ~=~ 4 (k^1k^2k^3)^{1/3} \,.
 \label{Radii-2}
 \end{equation}
Using the identity
\begin{equation}
c ~=~ \frac{a^2}{8} ~=~  \frac{Q_1 Q_5}{8 \kappa^2 R_y^2} ~=~ \frac{k^1 k^2 k^3}{\kappa R_y} \,,
 \end{equation}
the above change of coordinates can be written as
\begin{equation}
 \tau \;\equiv\;   \frac{t}{\kappa R_y} \;=\; \tilde{t} \,, \qquad 
\varphi ~\equiv~   \frac12 \psi -  \tilde{t}   \,, \qquad 
	\tilde{\phi} ~\equiv~ \phi   + \tilde{t} -  \varphi  \,.
 \label{coordsb-2}
 \end{equation}
%


\end{appendix}


\newpage

\begin{adjustwidth}{-3mm}{-3mm} 

\bibliographystyle{utphys}      

\bibliography{microstates}       

\end{adjustwidth}


\end{document}